\documentclass[12pt,epsf,amstex]{article}
\usepackage [dvips]{graphicx}
\usepackage{amsmath}
\usepackage{amssymb}
\usepackage{epsfig}

\addtocounter{secnumdepth}{1}
\newcounter{footnotestock}
\font\capfont=cmbx12 at 50 pt % or yinit, or...?
\newbox\capbox \newcount\capl \def\a{A}
\def\docappar{\medbreak\noindent\setbox\capbox\hbox{%
\capfont\a\hskip0.15em}\hangindent=\wd\capbox%
\capl=\ht\capbox\divide\capl by\baselineskip\advance\capl by1%
\hangafter=-\capl%
\hbox{\vbox to8pt{\hbox to0pt{\hss\box\capbox}\vss}}}
\def\cappar{\afterassignment\docappar\noexpand\let\a }
\begin{document}

\newcommand{\bP}{\bar{P}}
\newcommand{\br}{\bar{r}}
\newcommand{\bv}{\bar{v}}
\newcommand{\bbeta}{\bar{\beta}}
\newcommand{\bgamma}{\bar{\gamma}}
\newcommand{\btau}{\bar{\tau}}
\newcommand{\bomega}{\bar{\omega}}
\newcommand{\bOmega}{\bar{\Omega}}
\newcommand{\tbeta}{\tilde{\beta}}
\newcommand{\td}{\tilde{d}}
\newcommand{\tf}{\tilde{f}}
\newcommand{\tg}{\tilde{g}}
\newcommand{\tgamma}{\tilde{\gamma}}
\newcommand{\trho}{\tilde{\rho}}
\newcommand{\ttau}{\tilde{\tau}}
\newcommand{\tomega}{\tilde{\omega}}
\newcommand{\tN}{\tilde{N}}
\newcommand{\tT}{\tilde{T}}
\newcommand{\flead}{f_{\rm lead}}
\newcommand{\ffoll}{f_{\rm foll}}

\newcommand{\rholead}{{\rho}_{\rm lead}}
\newcommand{\rhoAlead}{{\rho}_{A,\rm lead}}
\newcommand{\rhoBlead}{{\rho}_{B,\rm lead}}
\newcommand{\rhoClead}{{\rho}_{C,\rm lead}}
\newcommand{\rhofoll}{{\rho}_{\rm foll}}
\newcommand{\rhoAfoll}{{\rho}_{A,\rm foll}}
\newcommand{\rhoBfoll}{{\rho}_{B,\rm foll}}
\newcommand{\sigmafoll}{{\sigma}_{\rm foll}}
\newcommand{\brho}{\bar{\rho}}
\newcommand{\brholead}{\bar{\rho}_{\rm lead}}
\newcommand{\brhofoll}{\bar{\rho}_{\rm foll}}
\newcommand{\brhoAlead}{\bar{\rho}_{A,\rm lead}}
\newcommand{\brhoBlead}{\bar{\rho}_{B,\rm lead}}
\newcommand{\brhoClead}{\bar{\rho}_{C,\rm lead}}
\newcommand{\brhoAfoll}{\bar{\rho}_{A,\rm foll}}
\newcommand{\brhoBfoll}{\bar{\rho}_{B,\rm foll}}
\newcommand{\brhoCfoll}{\bar{\rho}_{,\rm foll}}
\newcommand{\raAlead}{\ra_{A,\rm lead}}
\newcommand{\raBlead}{\ra_{B,\rm lead}}
\newcommand{\raClead}{\ra_{C,\rm lead}}
\newcommand{\pp}{{\prime\prime}}
\newcommand{\tdel}{t_{\rm del}}
\newcommand{\vm}{v_{\rm m}}

\newcommand{\hrho}{\hat{\rho}}
\newcommand{\htau}{\hat{\tau}}

\newcommand{\bE}{{\bf{E}}}
\newcommand{\bO}{{\bf{O}}}
\newcommand{\bR}{{\bf{R}}}
\newcommand{\bS}{{\bf{S}}}
\newcommand{\bT}{\mbox{\bf T}}
\newcommand{\bt}{\mbox{\bf t}}
\newcommand{\half}{\frac{1}{2}}
\newcommand{\summ}{\sum_{m=1}^n}
\newcommand{\sumq}{\sum_{q=1}^\infty}
\newcommand{\sumqno}{\sum_{q\neq 0}}
\newcommand{\prodm}{\prod_{m=1}^n}
\newcommand{\prodq}{\prod_{q=1}^\infty}
\newcommand{\maxm}{\max_{1\leq m\leq n}}
\newcommand{\minm}{\min_{1\leq m\leq n}}
\newcommand{\maxphi}{\max_{0\leq\phi\leq 2\pi}}
\newcommand{\bsA}{\mathbf{A}}
\newcommand{\bsV}{\mathbf{V}}
\newcommand{\bsE}{\mathbf{E}}
\newcommand{\bsT}{\mathbf{T}}
\newcommand{\bsZ}{\hat{\mathbf{Z}}}
\newcommand{\bse}{\mbox{\bf{1}}}
\newcommand{\bspsi}{\hat{\boldsymbol{\psi}}}
\newcommand{\cdottt}{\!\cdot\!}
\newcommand{\deltaR}{\delta\mspace{-1.5mu}R}
\newcommand{\invup}{\rule{0ex}{2ex}}

\newcommand{\bGamma}{\boldmath$\Gamma$\unboldmath}
\newcommand{\dd}{\mbox{d}}
\newcommand{\ee}{\mbox{e}}
\newcommand{\p}{\partial}
\newcommand{\expmVo}{\langle\ee^{-{\mathbb V}}\rangle_0}

\newcommand{\Rav}{R_{\rm av}}
\newcommand{\dav}{d_{\rm av}}
\newcommand{\Rc}{R_{\rm c}}
\newcommand{\rmax}{r_{\rm max}}

\newcommand{\la}{\langle}
\newcommand{\ra}{\rangle}
\newcommand{\rainr}{\rangle_r^{\rm in}}
\newcommand{\beq}{\begin{equation}}
\newcommand{\eeq}{\end{equation}}
\newcommand{\bea}{\begin{eqnarray}}
\newcommand{\eea}{\end{eqnarray}}
\def\lsim{\:\raisebox{-0.5ex}{$\stackrel{\textstyle<}{\sim}$}\:}
\def\gsim{\:\raisebox{-0.5ex}{$\stackrel{\textstyle>}{\sim}$}\:}

\numberwithin{equation}{section}

\thispagestyle{empty}

\title{\Large {\bf 
\vspace{-12mm}
Spontaneous symmetry breaking \\[2mm]
in a two-lane model \\[2mm]
for bidirectional overtaking traffic}}
%\phantom{xxx} }}
 
\author{{C. Appert-Rolland${}^{2,1}$, H.J. Hilhorst${}^{1,2}$, and
    G. Schehr${}^{2,1}$}\\[5mm] 
{\small ${}^{1}$ Univ. Paris-Sud, Laboratoire de Physique Th\'eorique,
  UMR8627}\\[-1mm]  
{\small B\^atiment 210, Orsay F-91405, France}\\
{\small ${}^{2}$ CNRS, Orsay F-91405, France}}

\maketitle

\begin{small}
\begin{abstract}
\noindent 
First we consider a unidirectional flux $\bomega$ 
of vehicles each of which is characterized by its `natural' 
velocity $v$ drawn from a distribution $P(v)$.
The traffic flow 
is modeled as a collection of straight
`world lines' in the time-space plane, with overtaking events 
represented by a fixed queuing time 
$\tau$ imposed on the overtaking vehicle.
This geometrical model exhibits platoon formation
and allows, among many other things, for the calculation of 
the effective average velocity $w\equiv\phi(v)$
of a vehicle of natural velocity $v$.
Secondly, we extend the model
to two opposite lanes, $A$ and $B$.
We argue that the queuing time $\tau$ in one lane
is determined by the traffic density in the
opposite lane. 
On the basis of reasonable additional assumptions 
we establish a set of
equations that couple the two lanes and can be solved
numerically. 
It appears that above a critical value $\bomega_{\rm c}$ of the
control parameter $\bomega$ the symmetry between the lanes is
spontaneously broken: there is 
a slow lane where long platoons form behind the slowest vehicles,
and a fast lane where overtaking is easy due to the wide spacing between
the platoons in the opposite direction.
A variant of the model is studied in which 
the spatial vehicle density $\brho$ rather than the flux $\bomega$ is the
control parameter. Unequal
fluxes $\bomega_A$ and $\bomega_B$ in the two lanes are also considered.
The symmetry breaking phenomenon exhibited by this model,
even though no doubt hard to observe in pure form in real-life traffic, 
nevertheless indicates a tendency of such traffic.
\vspace{3mm}

\noindent
{{\bf Keywords:} bidirectional traffic, overtaking traffic, geometric
  model, spontaneous symmetry breaking} 
\end{abstract}
\end{small}
\vspace{3mm}

\noindent LPT Orsay 10/38
\newpage

%%%%%%%%%%%%%%%%%%%%%%%%%%%%%%%%%%%%%%%%%%%%%%%%%%%%%%%%%%%%%%%%%%%%%%%%%%%%%
%%%%%%%%%%%%%%%%%%%%%%%%%%%%%%%%%%%%%%%%%%%%%%%%%%%%%%%%%%%%%%%%%%%%%%%%%%%%%
%%%%%%%%%%%%%%%%%%%%%%%%%%%%%%%%%%%%%%%%%%%%%%%%%%%%%%%%%%%%%%%%%%%%%%%%%%%%%

\section{Introduction} 
\label{secintroduction}

\cappar Traffic issues arise in many physical systems,
ranging from intracellular traffic to road traffic \cite{chowdhury_s_s00}.
The development of simple traffic models, such as exclusion processes,
has led to a deeper understanding of how such
out-of-equilibrium systems are driven
(for example how the introduction of a reaction time can induce
metastability \cite{appert_s01}).
Much effort has been devoted to unidirectional flows,
sometimes on several parallel tracks 
(or `lanes' in the road traffic vocabulary)
\cite{harris_s05,pronina_k04,pronina_k06,reichenbach_f_f06,reichenbach_f_f07,%
reichenbach_f_f08,schiffmann_a_s10}. 
However, many applications
involve bidirectional flows.
It is only recently that models for bidirectional traffic
have been proposed.

%%%%%%%%%%%%%%%%%%%%%%%%%%%%%%%%%%%%%%%%%%%%%%%%%%%%%%%%%%%%%%%%%%%%%%%%%%%%%%

\subsection{Bidirectional traffic and symmetry breaking}
\label{secbidirectional}

In one set of models 
the two lanes are represented by
a single chain of lattice sites. 
Oppositely moving entities (vehicles,
particles, molecular motors, ...)  share
the same lane and certain exchange rules have to be defined for when
they meet. Recently
such a model has been solved exactly for a whole distribution
of hopping rates \cite{vanbeijeren10}.
In a variant called the ``bridge model'', oppositely moving
particles have to slow down when they meet,
as on a narrow bridge
\cite{evans95a,godreche95,erickson05,willmann_s_g05,grosskinsky_s_w07,%
popkov_e_m08}.

In another set of models  
the two lanes are represented explicitly.
When both lanes are accessible to particles moving in either
direction,
jams may be formed \cite{Kornissetal99,georgiev_s_z05,georgiev_s_z06,%
ebbinghaus_s09}.
By contrast, the traffic flow is quite efficient when oppositely
moving particles are confined to different lanes \cite{klumpp_l04}.
Even in that case, however,
% it may be necessary to take into account
residual interactions between particles
on different lanes can be taken into account
(e.g. \cite{popkov_s03}). For example, in \cite{lee_p_k97}
a particle (or vehicle) has to slow down when 
it crosses another one on the other lane;
however, the authors restrict themselves to
the special case where one of the lanes contains only one vehicle.
The influence on two-lane bidirectional traffic
of quenched disorder (in the sense of statistical physics)
\cite{juhasz10}, or of the dynamics of the track itself 
(in the context of intracellular traffic) \cite{ebbinghaus_a_s10},
has also been considered.
All the models that we have mentioned so far are defined on
a discrete space and the motion consists of hopping events.
\vspace{5mm}

One phenomenon of interest in traffic models
is the spontaneous breaking of symmetry, whether it be between two
-- in principle equivalent -- lanes or between two flow directions.
Symmetry breaking in discrete traffic systems
has been reported 
for two-lane unidirectional traffic 
\cite{popkov_p00,popkov_p01,melbinger10}
and for one-lane bidirectional traffic (e.g. \cite{klumpp_l04}, or
in the bridge model \cite{evans95a,%
godreche95,erickson05,willmann_s_g05,grosskinsky_s_w07,popkov_e_m08}).
In \cite{Schmittmannetal05,pronina_k07,jiang07} 
symmetry breaking occurs for two-lane
bidirectional 
traffic, but the only interaction between the lanes is through
narrow road entrances -- which may be seen as localized bridges -- at the
two road ends.

\begin{figure}%[hbt]
\begin{center}
\scalebox{.55}
%{\includegraphics{primlanesa.eps}}
{\includegraphics{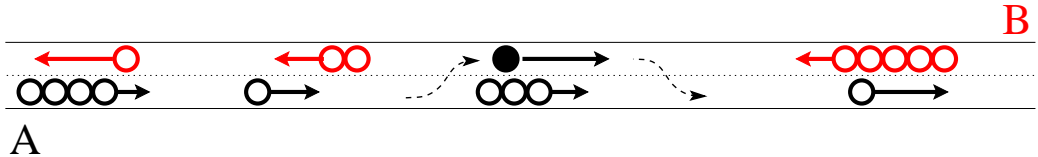}}
\end{center}
\caption{\small 
A two-lane road has vehicles (black circles)
on lane $A$ moving to the right and
vehicles (red circles) on lane $B$ moving to the left.
Arrows of variable length correspond to a distribution of velocities.
Strings of circles joined together indicate platoons
of faster vehicles queuing behind a slower one. In the center of the figure
a black vehicle (filled black dot) is overtaking a platoon moving
in the same direction. Platoon formation, overtaking events, and interaction
between the two lanes are the phenomena that we seek to describe 
in this work. In the actual model the vehicles and the platoons are 
point-like.}
\label{figsketch}
\end{figure}

In the present work our motivation is to construct a description 
of the traffic situation depicted in figure \ref{figsketch}
using a minimum number of mathematical ingredients.
This leads us to
propose a new two-lane model which is continuous in space and time.
It is simple enough to allow for 
theoretical analysis, yet retains the essentials of real-life traffic
for low to moderate vehicle densities.
A continuous distribution of velocities is considered.
On a given lane
vehicles with different velocities may overtake each other
provided there is enough space on the other lane.

%%%%%%%%%%%%%%%%%%%%%%%%%%%%%%%%
%%%%%%%%%%%%%%%%%%%%%%%%%%%%%%%%
%%%%%%%%%%%%%%%%%%%%%%%%%%%%%%%%
%%%%%%%%%%%%%%%%%%%%%%%%%%%%%%%%

The paper consists of two parts. In the first part, 
section \ref{secsinglelane},
we consider only a single lane of the two-lane road
and the influence of the other lane comes in only
through a postulated
input function.
In the second part, section \ref{secopposite},
we consider two opposite lanes coupled together and
determine the input function.
We report spontaneous symmetry breaking 
between the lanes
and are able to demonstrate
analytically how it results from the bulk evolution rules.
In section \ref{secnumerical} we show numerical results. In section
\ref{secmathematical} we collect a number of comments on the exact mathematical
status of the present one- and two-lane models.
Section \ref{secconclusion} is our conclusion.

%%%%%%%%%%%%%%%%%%%%%%%%%%%%%%%%%%%%%%%%%%%%%%%%%%%%%%%%%%%%%%%%%%%%%%%%%%%%%%

\subsection{The present model}
\label{secpresent}

%%%%%%%%%%%%%%%%%%%%%%%%%%%%%%%%%%%%%%%%%%%%%%%%%%%%%%%%%%%%%%%%%%%%%%%%%%%%%

\subsubsection{First part: single-lane model}

\begin{figure} %[hbt]
\begin{center}
\scalebox{.55}
%{\includegraphics{prim10d.eps}}
{\includegraphics{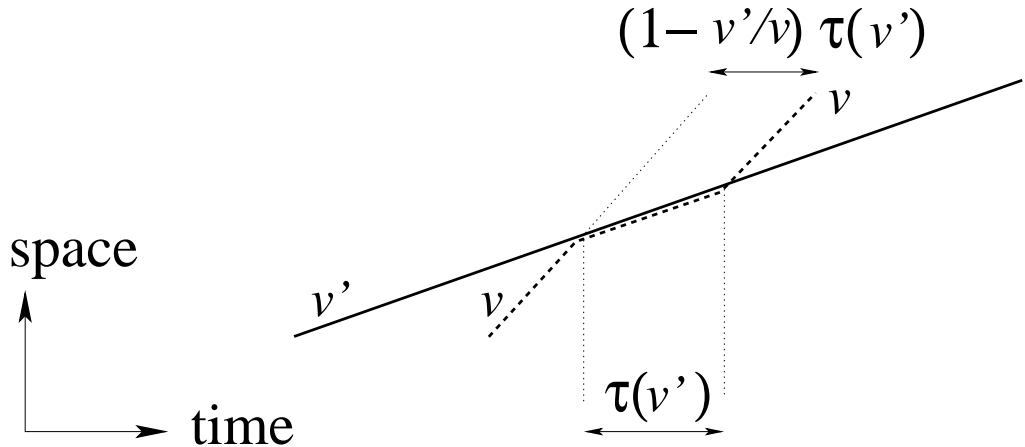}}
\end{center}
\caption{\small The basic overtaking event of a vehicle of velocity $v'$
by a vehicle of higher velocity $v$. During the time $\tau(v')$ necessary
for this event the two vehicles form a 2-platoon.
As a result the trajectory of the overtaking vehicle is displaced to
the right by an amount $\tdel(v,v')=(1-v'/v)\tau(v')$.}
\label{figprim1}
\end{figure}

The single-lane model of section \ref{secsinglelane} is purely geometric. 
It is governed by the rule (see figure \ref{figprim1}) that a
vehicle with velocity $v$ can overtake a slower one with 
velocity $v'<v$ only after it has stayed queuing behind that slower vehicle
during a {\it queuing time\,} time $\tau(v')$. Here $\tau(v')$ is 
the waiting time that elapses, on average, before the queuing
vehicle finds a sufficiently long free time interval to execute the 
overtaking maneuver. 
In the terminology of physics it is an interaction constant between the 
vehicles that depends 
only on the velocity of the slower vehicle.
A special case is obtained by taking the queuing time constant,
$\tau(v)=\btau$.
In the single-lane model $\tau(v')$ is an arbitrarily
postulated function.
Its expression 
should be derived, in principle, from the traffic density in the opposite
lane, which however at the level of the single-lane model is not described
explicitly. 

During the time interval $\tau(v')$ 
the vehicle of velocity $v'$ is the {\it leader\,}
and the vehicle of velocity $v$ is the {\it follower}.
After the time interval $\tau(v')$ the faster vehicle resumes its original,
or `natural' velocity $v$.
By a simple geometric argument in the time-space plane,
the vehicle with the higher velocity incurs a time delay
(a displacement to the right of its trajectory in the time-space plane) 
equal to%
%%%%%%%%%%
\footnote
{In functions with two velocity arguments, the first argument (velocity of the
vehicle that overtakes) will
always be larger than the second one (velocity of the vehicle being
overtaken).} 
\setcounter{footnotestock}{\thefootnote}
%%%%%%%%%%
\beq
\tdel(v,v') = \left( 1-\frac{v'}{v} \right)\tau(v').
\label{deftdel}
\eeq

%%%%%%%%%%%%%%%%%%%%%%%%%%%%%%%%
%%%%%%%%%%%%%%%%%%%%%%%%%%%%%%%%
%%%%%%%%%%%%%%%%%%%%%%%%%%%%%%%%
%%%%%%%%%%%%%%%%%%%%%%%%%%%%%%%%

At high traffic densities more than one vehicle may be queuing
simultaneously behind the same slower vehicle and hence
overtaking events will occur that involve 
three or more vehicles. 
The trajectories for such many-vehicle events may be constructed
with the aid of an appropriate interpretation of
the same queuing time rule (see section \ref{secsinglelane}).
The rules then allow for the construction of the full set of individual
vehicle trajectories in the time-space ($tx$) plane. 
We call this model `geometric' because
the resulting vehicle trajectories form a pattern of straight lines
(a sample pattern will be shown in section 
\ref{secsinglelane}, figure 5).

The construction shows that at any point of time each vehicle is either 
a follower (that is, it is being delayed behind another vehicle)
or, if not, is a leader and
advances at its natural velocity accompanied by any number
(possibly zero) of followers.
A leader accompanied by $n-1$ followers will be called an 
$n$-{\it platoon}; a $1$-platoon, therefore, is a single free vehicle.

In space-time diagrams like those of figure \ref{figprim1}
a moving platoon is represented by coinciding straight-line segments.
In our model both single vehicles and platoons of vehicles are
point-like: they have zero length in space. As a consequence, the model can
be expected to apply to low and moderate vehicle densities $\rho(v)$, 
but not to very high ones.
For this single-lane problem
we will determine, in particular, the two quantities most
characteristic of a stationary traffic flow. These are the spatial density
$\rho_{\rm lead}(v)$ of leaders (which is also the platoon density)
and the spatial density $\rho_{\rm foll}(v)$ of followers,
for a given velocity $v$. The expressions for these quantities
will depend on the queuing time
function $\tau(v)$ and on the boundary conditions. 
The latter may be imposed either at point of entry $x=0$
of the road section (for a road section with open-ended boundary
conditions), or at an initial time $t=0$ (for a ``ring'': a road section with
periodic boundary conditions); the two
possibilities give rise to slightly different mathematics.
\vspace{3mm}

A few words are in place here about other geometric single-lane models
occurring in the literature. As an example we cite
Ben-Naim, Krapivsky, and Redner \cite{BenNaimetal94},  
in whose model a vehicle or a platoon
that catches up with another one assumes that one's
velocity without ever overtaking it; the model
therefore describes clustering into platoons that can only grow.
It has no stationary state but is appropriate for the analysis of
platoon formation, for which the authors derive scaling laws.
In an extension of the same model
Ben-Naim and Krapivsky \cite{BenNaimKrapivsky97,BenNaimKrapivsky98}
consider overtaking based on vehicles having
an ``escape rate'' from the platoons; there is then a stationary state 
which may be analyzed by a Boltzmann type approach.

%%%%%%%%%%%%%%%%%%%%%%%%%%%%%%%%%%%%%%%%%%%%%%%%%%%%%%%%%%%%%%%%%%%%%%%%%%%%%

\subsubsection{Second part:  two-lane model}

In section \ref{secopposite} of this paper we consider
two opposite lanes called $A$ and $B$, as represented in figure
\ref{figsketch}, each one described
individually by the single-lane model of section \ref{secsinglelane}
with queuing times $\tau_A(v)$ and $\tau_B(v)$.
The queuing time $\tau_A(v)$ is supposed to have
its origin in the traffic density in lane $B$ and {\it vice versa}. 
On the basis of additional hypotheses we express $\tau_A(v)$ explicitly
in terms of the traffic density in lane $B$, and {\it vice versa}.
This essentially amounts, in physical parlance, 
to establishing a mean-field type interaction 
between the traffic flows on the two lanes.
It is again possible to study the resulting model analytically.
The simplest case occurs when there is full symmetry between the lanes 
$A$ and $B$.
Let a parameter $\bomega$ represent the intensity of the traffic flow.
The most interesting result is that in that case,
when $\bomega$ crosses a critical value $\bomega_{\rm c}$,
the symmetry between the lanes is spontaneously broken:
one lane, say $A$, will have a high density of short platoons
with on average a higher velocity than the second lane, which will have a
lower density of longer platoons.
The platoon density difference $\rhoAlead-\rhoBlead$ 
plays the role of the order parameter, as will be seen in
figure \ref{symdens}. 
We describe this phase transition
and extend our analysis and simulations to the non-symmetric case.

Because of the nonlocal (mean-field) character of the lane-lane coupling,
this two-lane model, contrary to the single-lane models on which it is
based, does not allow any longer for a geometric 
representation. That is, there is no corresponding model
of individually interacting vehicles.
In a companion paper \cite{appert_h_s10b}
we will present a truly microscopic geometric {\it two-}lane problem
and compare its simulation results to the theory of this work.

%%%%%%%%%%%%%%%%%%%%%%%%%%%%%%%%%%%%%%%%%%%%%%%%%%%%%%%%%%%%%%%%%%%%%%%%%%%%%%

\section {A single-lane model}
\label{secsinglelane}

We will first complete the definition of the single-lane model
described in section \ref{secpresent}. This model has 
unidirectional traffic and a basic overtaking event
as pictured in figure \ref{figprim1}.
The delay time $\tau(v)$ is a fixed given function.
We will associate with each natural velocity $v$ an effective or
average velocity $w=\phi(v)$, then find an equation for
the function $\phi$,
and show how the platoon statistics may be derived from its solution.

%%%%%%%%%%%%%%%%%%%%%%%%%%%%%%%%%%%%%%%%%%%%%%%%%%%%%%%%%%%%%%%%%%%%%%%%%%%%

\subsection{Preliminaries}
\label{secpreliminaries}

%%%%%%%%%%%%%%%%%%%%%%%%%%%%%%%%%%%%%%%%%%%%%%%%%%%%%%%%%%%%%%%%%%%%%%%%%%%%

\subsubsection{Many-vehicle events}
\label{secmanycarevents}

\begin{figure}
\begin{center}
\scalebox{.55}
%{\includegraphics{prim21f.eps}}
{\includegraphics{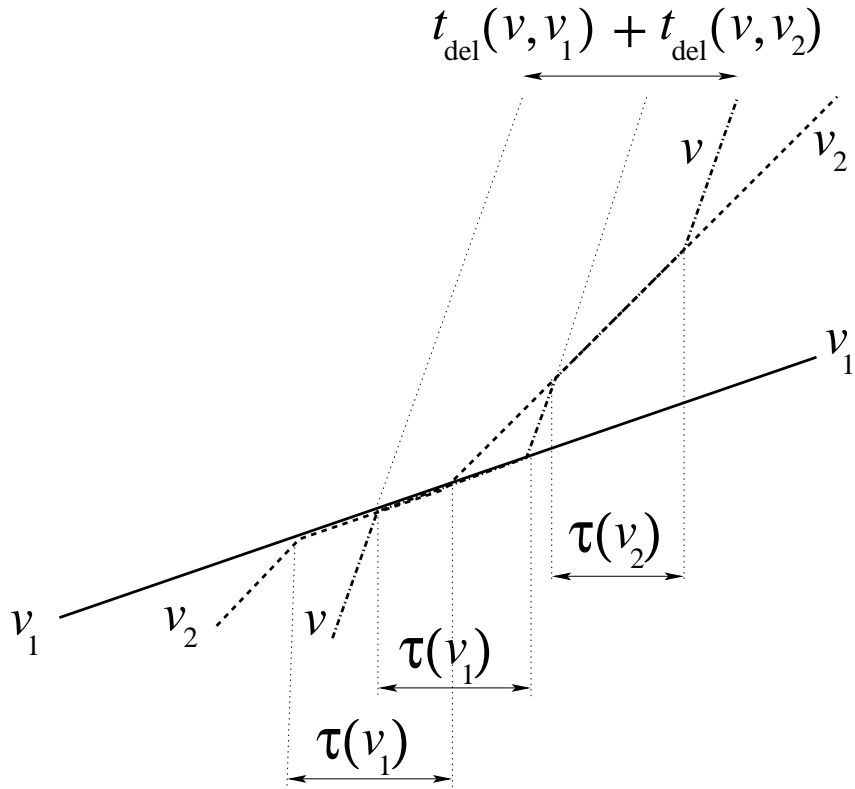}}
\end{center}
\caption{\small Example of a three-vehicle overtaking event
with the time and space axes as in figure \ref{figprim1}.
First, the vehicle of velocity $v_1$ is overtaken by those of velocities
$v_2$ and $v$ during partially overlapping time intervals both of
length $\tau(v_1)$.
During the overlapping part of those intervals there is a 3-platoon. 
Next, in a separate event of duration $\tau(v_2)$,
the vehicle of velocity $v_2$ is overtaken by the one of velocity $v$.
As a result the latter has incurred a time delay 
$\tdel(v,v_1)+\tdel(v,v_2)$.}
\label{figprim2}
\end{figure}

We must begin by briefly dwelling 
on overtaking events involving three or more vehicles.
For these we will adopt the rule that queuing times, and hence 
incurred delays, are additive.
Figure \ref{figprim2} shows an example of a three-vehicle event.
The vehicle of velocity $v$
overtakes successively two slower
vehicles of velocities $v_1$ and $v_2$ 
and as a result incurs a time delay 
$\tdel^{\rm tot}=\tdel(v,v_1)+\tdel(v,v_2)$ with $\tdel$ given by
(\ref{deftdel}). 
One may note that the vehicle of velocity $v_2$ also overtakes the one of
velocity $v_1$ and that there is a time interval during which the latter
has two followers (there is a 3-platoon).

\begin{figure}
\begin{center}
\scalebox{.65}
{\includegraphics{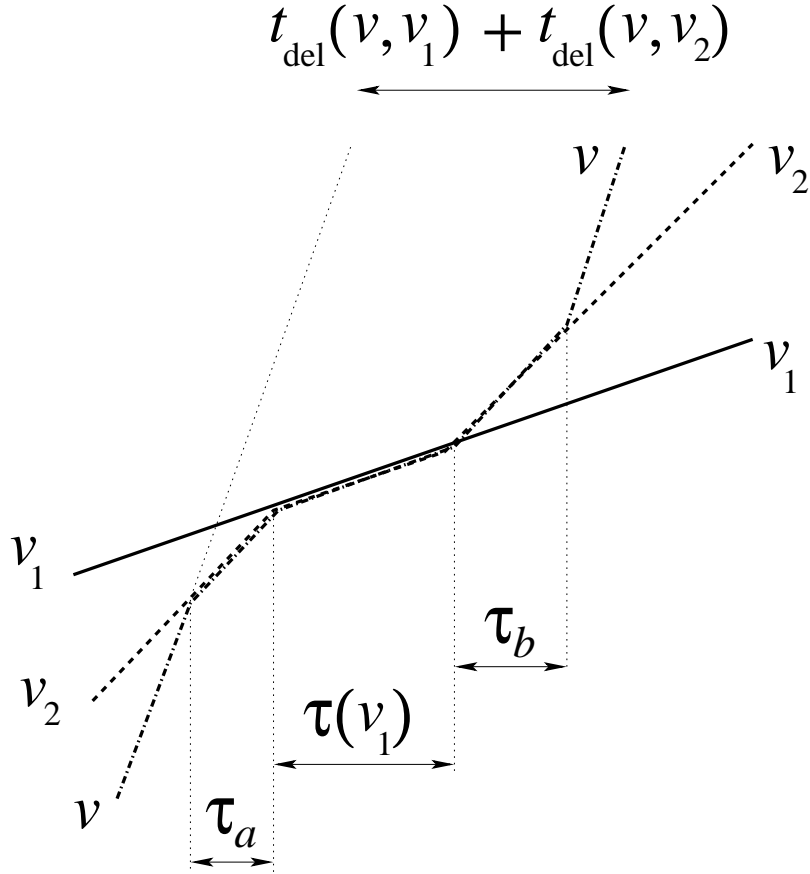}}
\end{center}
\caption{\small A different example of a three-vehicle overtaking
  event; the time and space axes are as in figure \ref{figprim1}.
When the vehicle of velocity $v$ has queued for a time $\tau_a$ 
behind the one of velocity $v_2$, the two catch up
with the still slower vehicle of velocity $v_1$. During
a time interval $\tau(v_1)$
the two simultaneously spend the required queuing time $\tau(v_1)$
behind that vehicle, which they then simultaneously overtake.
Following that, the vehicle of velocity $v$ spends
the remaining time $\tau_b=\tau(v_2)-\tau_a$ queuing behind the one of 
velocity $v_2$, after which it overtakes it.
The net result is that the vehicle of velocity $v$ 
has incurred the same time delay 
$\tdel(v,v_1)+\tdel(v,v_2)$ as in figure \ref{figprim2}.}
\label{figprim3}
\end{figure}

In  figure \ref{figprim3} a slightly more complicated three-vehicle overtaking 
event is depicted.
The vehicle of velocity $v$ 
spends two disjoint time intervals $\tau_a$ and $\tau_b$ queuing
behind the one of velocity $v_2$, but such that $\tau_a+\tau_b=\tau(v_2)$,
thus satisfying the rule imposing the total queuing time.

\begin{figure}
\begin{center}
\scalebox{.55}
%{\includegraphics{primkeventb.eps}}
{\includegraphics{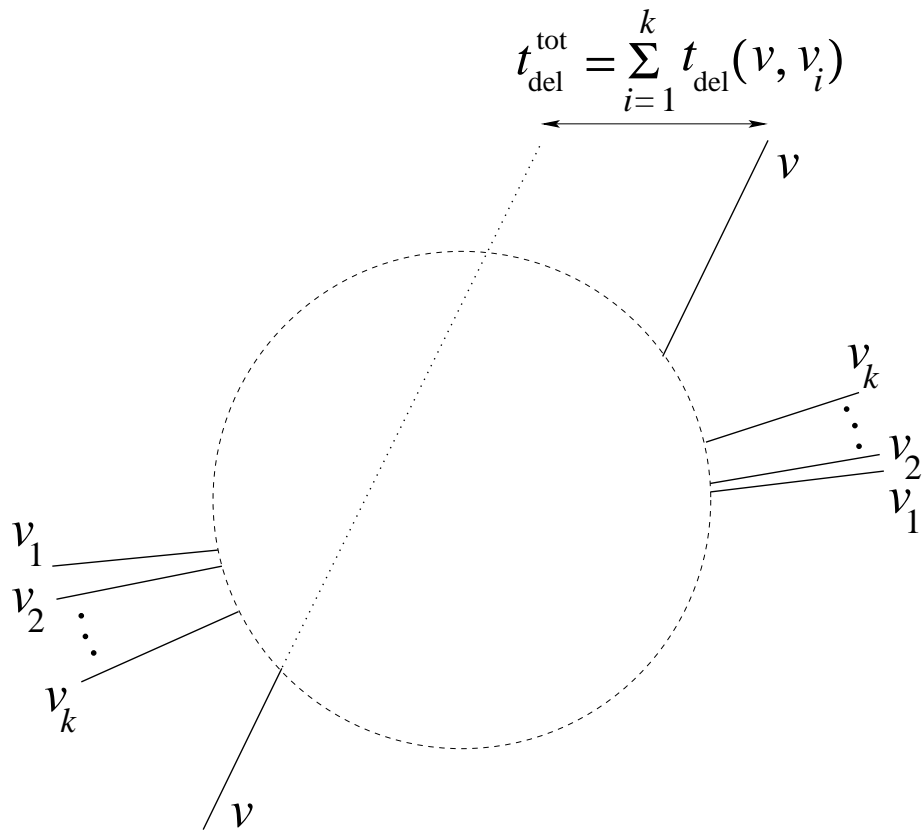}}
\end{center}
\caption{\small A vehicle of velocity $v$ overtakes $k$ slower vehicles. Its
  final trajectory suffers a time delay $t_{\rm del}^{\rm tot}$ which
  is the sum of the time delays incurred due to $k$ individual 
  overtaking events. The full sequence of partially overlapping
  overtakings is a complicated event that occurs inside the
  dashed disk. It has not been presented explicitly: only the final outcome
  is relevant in our description.}
\label{figprim4}
\end{figure}

One may construct arbitrary many-vehicle events respecting the same rule
by drawing the trajectories of vehicles of successively higher velocities.
As a consequence, in an event where
a vehicle of velocity $v$ overtakes $k$ other vehicles, its
outgoing trajectory in the $tx$ plane
will again be a straight line displaced by an amount 
$\tdel^{\rm tot}=\sum_{i=1}^k\tdel(v,v_i)$
to the right with respect to its incoming trajectory.
This is exhibited in figure \ref{figprim4}.

\begin{figure}
\begin{center}
\scalebox{.50}
%{\includegraphics{primtraja.eps}}
{\includegraphics{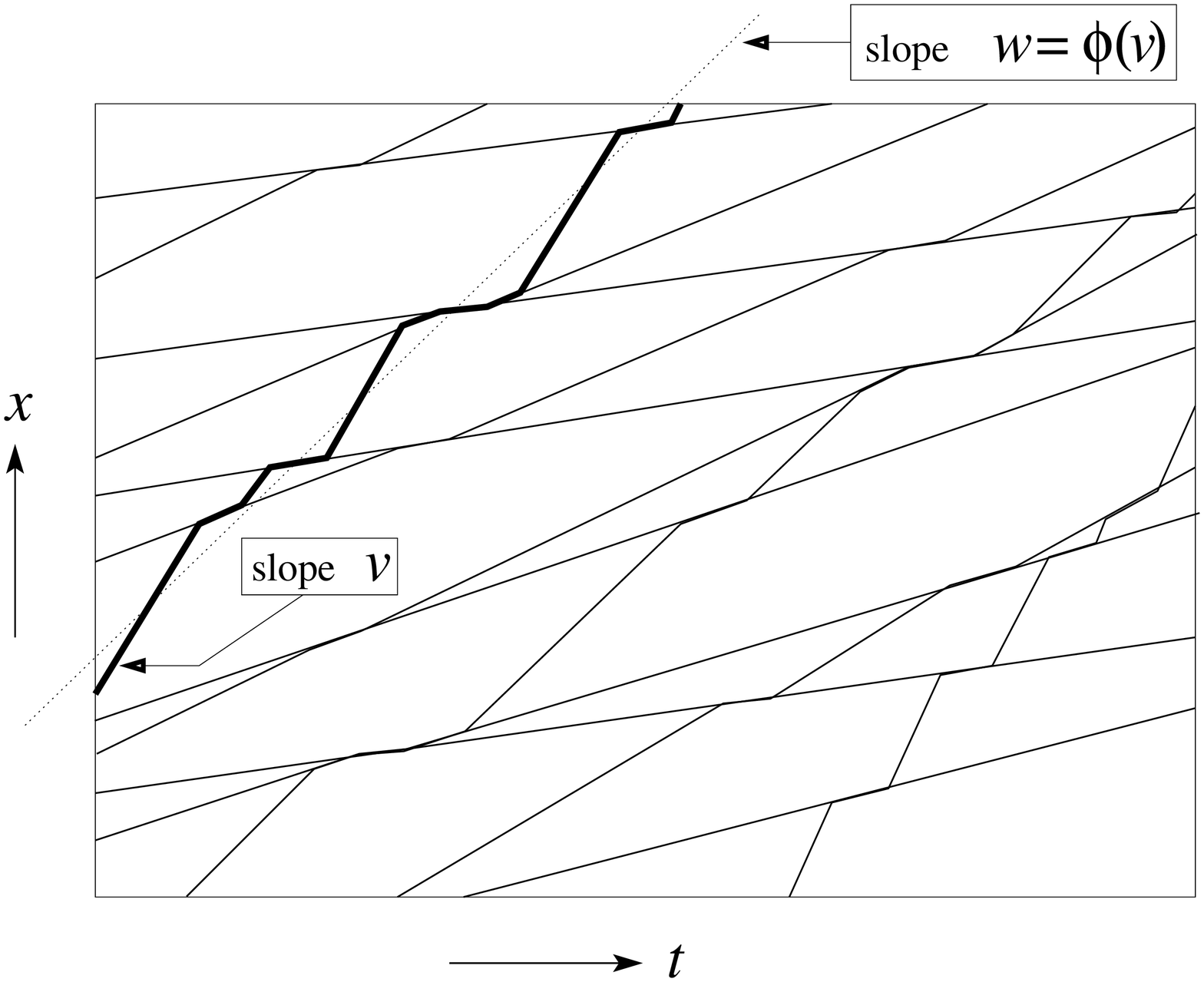}}
\end{center}
\caption{\small Sample pattern of vehicle trajectories in the $tx$
  plane. The heavy line is the trajectory of a vehicle of natural
  velocity $v$. When averaged over a long time interval, it appears to
  have an effective velocity equal to the slope of the dotted line and
  denoted by $w=\phi(v)$.}
\label{figprim5}
\end{figure}

In figure \ref{figprim5}, finally, we show a sample set of trajectories
on a given lane section during a given time interval.

%%%%%%%%%%%%%%%%%%%%%%%%%%%%%%%%
%%%%%%%%%%%%%%%%%%%%%%%%%%%%%%%%
%%%%%%%%%%%%%%%%%%%%%%%%%%%%%%%%
%%%%%%%%%%%%%%%%%%%%%%%%%%%%%%%%

%%%%%%%%%%%%%%%%%%%%%%%%%%%%%%%%
%%%%%%%%%%%%%%%%%%%%%%%%%%%%%%%%
%%%%%%%%%%%%%%%%%%%%%%%%%%%%%%%%
%%%%%%%%%%%%%%%%%%%%%%%%%%%%%%%%

%%%%%%%%%%%%%%%%%%%%%%%%%%%%%%%%
%%%%%%%%%%%%%%%%%%%%%%%%%%%%%%%%
%%%%%%%%%%%%%%%%%%%%%%%%%%%%%%%%
%%%%%%%%%%%%%%%%%%%%%%%%%%%%%%%%

%%%%%%%%%%%%%%%%%%%%%%%%%%%%%%%%
%%%%%%%%%%%%%%%%%%%%%%%%%%%%%%%%
%%%%%%%%%%%%%%%%%%%%%%%%%%%%%%%%
%%%%%%%%%%%%%%%%%%%%%%%%%%%%%%%%

%%%%%%%%%%%%%%%%%%%%%%%%%%%%%%%%%%%%%%%%%%%%%%%%%%%%%%%%%%%%%%%%%%%%%%%%%%%%

\subsubsection{Boundary conditions}
\label{secboundaryconditions}

In order for the single-lane problem to be well defined it has to be
supplemented with boundary conditions.
In the $tx$ plane these may be applied either along the space axis or
along the time axis. 
Both possibilities have merits, as we will
briefly discuss now.
\vspace{3mm}

{\it Boundary condition in $x=0$ for all times (open-ended boundary
  conditions).\,\,--} 
One may consider an open-ended lane section of coordinates $0\leq x\leq L$
and impose, for each $v$, the rate $\omega(v)dv$
at which vehicles with natural velocity between $v$ and $v+\dd v$
enter the lane section at the point $x=0$.
We will set
\beq
\bomega = \int\!\dd v\,\omega(v), \qquad \omega(v)=\bomega P(v),
\label{defOBC}
\eeq
where now $P(v)\dd v$ is the probability that an entering vehicle
picked at random has a velocity in $[v,v+\dd v]$.
In $x=L$ the vehicles leave the lane section freely.
We will refer to this situation as having {\it open-ended\,} boundary
conditions (OBC). 
Because of the continuity equation,
in a stationary state all vehicles that enter the lane at $x=0$
at some rate must also cross
any other observation point $x>0$ at the same rate.
Hence the control function $\omega(v)$, once given for $x=0$,
will be the same at any other point in space.
\vspace{3mm}

{\it Boundary conditions at $t=0$ in all of space (periodic boundary
  conditions).\,\,--} 
One may consider, especially in the context
of simulation \cite{appert_h_s10b} or of theoretical work,
a circular lane of circumference $L$.
It is then natural to
impose at an initial time $t=0$ the spatial density $\rho(v)\dd v$
of vehicles with velocity between $v$ and $v+\dd v$.
We will set
\beq
\brho=\int\!\dd v\,\rho(v), \qquad \rho(v)=\brho R(v),
\label{defPBC}
\eeq
where now $R(v)\dd v$ is the probability that  
a vehicle picked at random
at the initial time has a velocity in $[v,v+\dd v]$.
We will refer to this situation as having {\it periodic\,} boundary
conditions (PBC).
Since vehicles cannot leave or enter the ring, the control function
$\rho(v)$, when given at some initial time $t=0$, 
will be the same at all later times $t>0$.
 
We will refer to the probability densities
$P(v)$ and $R(v)$ as the `control functions', and to the parameters
$\bomega$ and $\brho$ that multiply them as the `control parameters'.
As shown above, both control functions are actually bulk quantities.
\vspace{3mm}

The boundary conditions as we have formulated them, 
whether at $x=0$ or at $t=0$, do not specify any correlations, 
or the absence thereof, that may
exist between the positions and/or the velocities of the 
vehicles on these boundaries. We will briefly return to this point in section
\ref{secmathematical}.
In any case, we will make the `physically reasonable' assumption
that there is a penetration length 
$x_{\rm  pen}$ in the case of OBC
(and a relaxation time $t_{\rm rel}$ in the case of PBC)
such that for $x\gg x_{\rm pen}$ (for $t\gg t_{\rm rel}$)
a state is reached that is both homogeneous in space
and stationary in time.
%\footnote%
%%%%%%%%%%%%%
%{In fact, we suppose tacitly, in addition, that any effect of the boundary
%  correlations that were to survive in the stationary state, would
%  have no influence on the results that we will derive below.}.
%%%%%%%%%%%%%
We will not attempt here to express $x_{\rm pen}$
in terms of $\omega(v)$ and $\tau(v)$, or $t_{\rm rel}$
in terms of $\rho(v)$ and $\tau(v)$.

%%%%%%%%%%%%%%%%%%%%%%%%%%%%%%%%%%%%%%%%%%%%%%%%%%%%%%%%%%%%%%%%%%%%%%%%%%%%%

\subsubsection{Natural velocity $v$ and effective velocity $w$}
\label{secnaturaleffective}

The first question that comes to mind is suggested by figure \ref{figprim5}. 
Suppose the traffic is in a stationary state.
What will then be the effective velocity $w$, 
that is, the velocity averaged over a sufficiently long time, 
of a vehicle that has $v$ as its natural velocity?
Let us assume that there exists a relation yielding $w$ as a function
of $v$ and write it as 
\beq
w=\phi(v), \qquad v=\psi(w),
\label{defphipsi}
\eeq 
where $\psi$ is the inverse of $\phi$.
We would like to determine $\phi(v)$ for all $v>v_{\rm 0}$.
Let us denote the minimum velocity 
(which may be zero) by $v_{\rm 0}$\,, that is,
$P(v)=0$ (for OBC) or $R(v)=0$ (for PBC) when $v < v_{\rm 0}$.
The vehicles with $v=v_{\rm 0}$ follow unperturbed trajectories: 
since they cannot overtake any other vehicle,
they have straight world lines in the $tx$ plane.
We therefore have $w_{\rm 0}\equiv\phi(v_{\rm 0})=v_{\rm 0}$.
Furthermore $\phi(v)$ should increase with $v$.

Below we will find the expression for $\phi(v)$ and see that it is
boundary condition independent.
In subsection \ref{secopenended} we will discuss the case of OBC;
in section \ref{seccircular}
we will then describe succinctly what changes in the case of PBC.\\

%%%%%%%%%%%%%%%%%%%%%%%%%%%%%%%%%%%%%%%%%%%%%%%%%%%%%%%%%%%%%%%%%%%%%%%%%%%%

\subsection{Open-ended traffic lane}
\label{secopenended}

%%%%%%%%%%%%%%%%%%%%%%%%%%%%%%%%%%%%%%%%%%%%%%%%%%%%%%%%%%%%%%%%%%%%%%%%%%%%

\subsubsection{General}
\label{secOBCgeneral}

Without the delays due to overtakings (that is: for $\tau(v)=0$, 
or equivalently: 
in the absence of `interaction' between the vehicles) 
the spatial density $\rho^{(0)}(v)\dd v$ of vehicles of natural
velocity in $[v,v+\dd v]$ is given by
\beq
\rho^{(0)}(v)\dd v = \frac{\omega(v)}{v}\dd v. 
\label{relrhoalpha}
\eeq
The factor $v^{-1}$ here indicates 
that a fast vehicle spends less time in the lane section and therefore
contributes less to the density.
The total vehicle density, again for $\tau(v)=0$, is therefore
\beq
\brho^{(0)}=\int_{v_0}^\infty \!\dd v\, \rho^{(0)}(v)
=\int_{v_0}^\infty\!\dd v\,\frac{\omega(v)}{v}\,.
\label{rhotot}
\eeq

This work is concerned with the modifications that occur due to the
interaction between the vehicles.

In the stationary state, at any fixed observation point 
vehicles of effective velocity $w$ will pass
with a probability of ${\tomega}(w)$ per unit of time.
We may find ${\tomega}(w)$ by first noting that
\beq
\omega(v)\dd v = {\tomega}(w)\dd w,
\label{relalphhalph}
\eeq
which is the continuity equation.
In (\ref{relalphhalph}) and henceforth, tildes will indicate quantities
expressed as functions of the effective velocity $w$.
It follows from (\ref{relalphhalph}) and (\ref{defphipsi}) that
\beq
{\tomega}(w) = \omega\big(\psi(w)\big)\psi'(w).
\label{extalpha}
\eeq
The spatial density ${\trho}(w)$
of vehicles with effective velocity in $[w,w+\dd w]$ is
\beq
{\trho}(w)\dd w = \frac{\tomega(w)}{w}\dd w  
= \frac{\omega(v)}{\phi(v)}\dd v. 
\label{exhrhow}
\eeq
The spatial density  $\rho(v)\dd v$ of
vehicles of natural velocity in $[v,v+\dd v]$ is defined by
\beq
\rho(v)\dd v = \trho(w)\dd w.
\label{deftrhov}
\eeq
We recall that each vehicle has a natural velocity that is fixed once
and for all. 
The density $\rho(v)$ takes into account all vehicles of natural velocity $v$
regardless of their actual velocity at the moment.
It thus appears from (\ref{exhrhow}) and (\ref{deftrhov}) that 
\beq
\rho(v) = \frac{\omega(v)}{\phi(v)}\,. 
\label{extrhov}
\eeq
This is a key equation of the present model, valid in its stationary state.

By integrating (\ref{extrhov}) over $v$ we see that
the total vehicle density in the interacting system 
is equal to
\beq
\brho = \int_{v_0}^\infty\!\dd v\,\frac{\omega(v)}{\phi(v)}\,.
\label{extrhotot}
\eeq
At this point $\phi(v)$ is still unknown.

%%%%%%%%%%%%%%%%%%%%%%%%%%%%%%%%%%%%%%%%%%%%%%%%%%%%%%%%%%%%%%%%%%%%%%%%%%%%%%

\subsubsection{Deriving an equation for $\phi(v)$}
\label{seceqnphi}

We will now derive an equation from which $\phi(v)$ can be solved.
The basic idea is provided by the fact that
the trajectory of a vehicle with a given velocity is affected 
(`renormalized') only
by those of vehicles with lower velocities.
This idea can be exploited as follows.

We continue to 
suppose the system is in a stationary state and consider it during
a time interval $[0,T]$. In the limit of large $T$ all vehicles will have
well-defined effective velocities.
Consider a marked vehicle of natural velocity $v$, and hence of
effective velocity $w=\phi(v)$, 
starting in the origin $x=0$ at time $t=0$.
During the interval $[0,T]$ any vehicle with effective velocity $w'<w$ 
initially present
in the spatial domain $[0,(w-w')T]$ will be overtaken by the marked vehicle.
Let $\tN(w,w')\dd w'$ be the number of vehicles with velocity in 
$[w',w'+\dd w']$ overtaken by the marked vehicle (see footnote
$^{\thefootnotestock}$). 
In the limit of large $T$, again, this number will be given by
\beq
\tN(w,w')\dd w' = (w-w')T \times {\trho}(w')\dd w'.
\label{defNw}
\eeq 
We will write $\ttau(w)\equiv\tau(\psi(w))$ for the queuing time
necessary for overtaking a vehicle of velocity $w$.
The total duration of the time intervals occupied by
the overtaking events (\ref{defNw})
amounts to a time $\tT_{\rm foll}(w,w')\dd w'$ given by
\beq
\tT_{\rm foll}(w,w')\dd w' = \ttau(w') \tN(w,w')\dd w'.
\label{Tfollwwp}
\eeq
During this time the marked vehicle is a follower traveling only at velocity
$\psi(w')$ and hence covers a distance $\td_{\rm foll}(w,w')\dd w'$ given by
\beq
\td_{\rm foll}(w,w') \dd w' = \ttau(w') \tN(w,w')\psi(w')\dd w'.
\label{dfollwwp}
\eeq
By integrating 
this last expression over $w'$ and using (\ref{defNw}) we obtain the sum 
$\td_{\rm foll}(w)$ of the distances traveled by the marked vehicle during its
overtaking events, 
\beq
\td_{\rm foll}(w)= T \int_{w_{\rm 0}}^{w}\!\!\dd w'\,(w-w')\,
{\trho}(w')\psi(w')\ttau(w').
\label{defdover}
\eeq
The remaining time $T_{\rm lead}$, not used for overtaking events, is
\beq
\tT_{\rm lead}(w) = T - 
T \int_{w_{\rm 0}}^{w}\!\!\dd w'\,(w-w')\,{\trho}(w')\ttau(w').
\label{defTfree}
\eeq
During the time $\tT_{\rm lead}(w)$
the marked vehicle drove at its natural velocity $v$ and covered
a distance $\td_{\rm lead}(w)=v\tT_{\rm lead}(w)$, where our notation
is hybrid.
We now obtain the effective velocity $w$ of the marked vehicle as
\bea
w &=& (\td_{\rm lead}(w)+\td_{\rm foll}(w))/{T} \nonumber\\[2mm]
  &=& v - v \int_{w_{\rm 0}}^{w}\!\!\dd w'\,(w-w')\,{\trho}(w')\,\ttau(w')
        + \int_{w_{\rm 0}}^{w}\!\!\dd w'\,(w-w')\,{\trho}(w')
                                        \psi(w')\,\ttau(w'). \nonumber\\
&&
\label{exw}
\eea
Employing (\ref{defphipsi}), (\ref{extrhov}), and (\ref{extalpha})
in (\ref{exw}) and eliminating the variables $w$ and $w'$,
we reexpress (\ref{exw}) as 
\beq
\phi(v)\,=\,v\,-\,\int_{v_{\rm 0}}^v\!\dd v'\,\omega(v')\,\tau(v')
(v-v')\left[ \frac{\phi(v)}{\phi(v')}-1 \right],
\label{eqnphiv0}
\eeq
which may be rewritten as
\beq
\phi(v) = \frac{v+\int_{v_0}^v\!\dd v'\,\omega(v')\tau(v')(v-v')}
{1+\int_{v_0}^v\!\dd v'\,\omega(v')\tau(v')(v-v')[\phi(v')]^{-1}}\,.
\label{eqnphiv}
\eeq
This is the equation for $\phi(v)$ that we sought.
It depends on the shapes of the velocity distribution $\omega(v)$ and 
the queuing time distribution $\tau(v)$.

%%%%%%%%%%%%%%%%%%%%%%%%%%%%%%%%%%%%%%%%%%%%%%%%%%%%%%%%%%%%%%%%%%%%%%%%%%%%%%

\subsubsection{Solving the equation for $\phi(v)$}
\label{secsolvingphi}

The equation for $\phi(v)$ may be readily solved.
The easiest way to do this is to 
observe that it may be written as a linear equation in terms of $1/\phi(v)$.
In appendix
\ref{secappendixchi} it is shown that
\beq
\frac{1}{\phi(v)}=\frac{1}{v_0}\,-\,\int_{v_0}^v\!\dd v'\,\left[
v'\,+\,\int_{v_0}^{v'}\!\dd v^{{\pp}}\,(v'-v^{{\pp}})
\omega(v^{{\pp}})\tau(v^{{\pp}})
\right]^{-2}.
\label{solchiv}
\eeq 
One easily checks that $\phi(v_0)=v_0$ as had to be expected.
One also verifies that $\tau(v^{\pp})=0$ leads to $\phi(v)=v$,
as had to be the case, and that $\tau(v^{\pp})=\infty$ gives
$\phi(v)=v_0$, which is intuitively obvious.
\vspace{3mm}

There are a few cases in which the integrals in (\ref{solchiv}) can be carried
out exactly. One of them is the case where $\tau(v)=\btau$ is a constant and
$\omega(v)$ is a block distribution, constant for $v_0<v<v_{\rm m}$
and zero elsewhere. We present it as an example in appendix
\ref{secappexactsol}. 
However, we will not pursue such special cases here.
In the numerical section \ref{secnumerical} we present
a figure showing the typical behavior of $\phi(v)$ for two different
traffic fluxes.
\vspace{3mm}

In cases where the velocity distribution $\omega(v)$ is a sum of delta peaks,
the reasoning performed above leads to discretized versions of
equations (\ref{eqnphiv}), (\ref{solchiv}), 
and other results below. We briefly present these
in appendix \ref{secappdiscr}.

%%%%%%%%%%%%%%%%%%%%%%%%%%%%%%%%%%%%%%%%%%%%%%%%%%%%%%%%%%%%%%%%%%%%%%%%%%%%%%

\subsubsection{Platoons, leaders, and followers}
\label{secplatoons}

An $n$-platoon consists of $n-1$ vehicles all blocked behind
a single `leader' advancing at its natural velocity, say $v'$.
The $n-1$ `followers', although having natural velocities larger than $v'$, 
are also advancing at the velocity $v'$ of the leader.
Each follower is in the queuing interval $\tau$
after which it will overtake the leader.
It will be convenient to refer to $n$ as the `platoon length', 
even if in the present model platoons are point-like.
A freely advancing single vehicle without
followers will be called a `$1$-platoon'.
In this subsection we study the spatial density of platoons 
characterized by a given velocity and/or platoon length. 
Our approach will be to first study the statistics of the followers
and then derive from it the statistics of the leaders.
\vspace{5mm}

%%%%%%%%%%%%%%%%%%%%%%%%%%%%%%%%%%%%%%%%%%%%%%%%%%%%%%%%%%%%%%%%%%%%%%%%%%%%%%

{\it Followers. --\,\,}
Let $\ffoll(v)$ be the fraction of its time that a vehicle of natural velocity 
$v=\psi(w)$ spends following slower vehicles.
From equation (\ref{defTfree}) we see that
\bea
\ffoll(v) &=& \int_{w_0}^w \!\dd w'(w-w')\trho(w')\ttau(w') 
\nonumber\\[2mm]
&=& \int_{v_0}^v\!\dd v' \left[ \frac{\phi(v)}{\phi(v')}-1 \right]
\omega(v')\tau(v').
\label{exfv}
\eea
Similar reasoning, but without integrating on $v'$, 
shows that
(see footnote $^{\thefootnotestock}$)
\beq
f(v,v')\dd v' = 
\left[\frac{ \phi(v)}{\phi(v')}-1 \right] \omega(v')\tau(v')\dd v',
\qquad v>v',
\label{exfvvp}
\eeq
is the fraction of its time that a vehicle of natural velocity $v$ spends
following vehicles with velocities in $[v',v'+\dd v']$.
We will now pass from these time fractions to spatial densities.

Equation (\ref{extrhov}) gives the density $\rho(v)$ of vehicles of natural
velocity $v$. It follows that in a stationary state
the density of {\it following\,} vehicles of natural velocity $v$,
to be called $\rho_{\rm foll}(v)$, is given by 
\beq
\rho_{\rm foll}(v)=\rho(v)\ffoll(v).
\label{extrhopm}
\eeq
Hence, using (\ref{extrhov}) and (\ref{exfv}) in (\ref{extrhopm}) we obtain
\beq
\rho_{\rm foll}(v) 
= \frac{\omega(v)}{\phi(v)}
\int_{v_0}^v \!\dd v'
\left[\frac{\phi(v)}{\phi(v')}-1 \right] \omega(v')\tau(v').
\label{extrhovm}
\eeq
Let 
$\brho_{\rm foll}$ denote 
the total density of following vehicles.
This quantity is obtained as the integral on $v$ of equation
(\ref{extrhovm}). 
Hence
\beq
\brhofoll= \int_{v_0}^\infty \!\dd v \frac{\omega(v)}{\phi(v)}
\int_{v_0}^v \!\dd v'
\left[\frac{\phi(v)}{\phi(v')}-1 \right] \omega(v')\tau(v'). 
\label{ectrhototm}
\eeq
Let us define, for all $v>v'$,
\bea
\rho(v,v')\,\dd v\,\dd v' 
&=& \mbox{spatial density of vehicles of natural 
          velocity in $[v,v+\dd v]$} \nonumber\\
& & \mbox{following a leader of velocity in $[v',v'+\dd v']$}.
\eea
By interpreting (\ref{ectrhototm}) we see that
\beq
\rho(v,v')
= \frac{\omega(v)}{\phi(v)}\, \frac{\omega(v')}{\phi(v')}\,\tau(v')\,
    [\phi(v)-\phi(v')], \qquad v>v'.
\label{deftrhominvvp} 
\eeq
The two-velocity density function $\rho(v,v')$ together with the expression for
$\phi(v)$ are two fundamental quantities in the present model.
We have the obvious relation
\beq
\rhofoll(v) = \!\int_{v_0}^v\!\dd v'\, \rho(v,v').
\label{intrhofoll}
\eeq
In addition to (\ref{intrhofoll}) there is a second integral that we can
construct by projecting $\rho(v,v')$ onto the $v'$ axis,
\bea
{\sigmafoll}(v') = \int_{v'}^\infty\!\dd v\,\rho(v,v').
\label{intsigmafoll}
\eea
It has the interpretation of the total density of vehicles that follow a leader
of natural velocity $v'$.
We observe that
\beq
\brhofoll\,=\,
\int_{v_0}^\infty\!\dd v\,\rhofoll(v)=
\int_{v_0}^\infty\!\dd v'\,{\sigmafoll}(v'),
\label{sigmarho}
\eeq
which we record for later use.
\vspace{5mm}

%%%%%%%%%%%%%%%%%%%%%%%%%%%%%%%%%%%%%%%%%%%%%%%%%%%%%%%%%%%%%%%%%%%%%%%%%%%%%%

{\it Leaders. --\,\,}
The quantity $\flead(v) \equiv 1-\ffoll(v)$ is the fraction of its
time that a vehicle of natural velocity $v$ spends {\it not\,}
queuing: it may either constitute a $1$-platoon all by itself or be at
the head of an $n$-platoon. 
Upon setting correspondingly
\beq
\rholead(v) = \rho(v)-\rhofoll(v), 
\label{sumrulerho}
\eeq
we find with the aid of (\ref{extrhovm}) that 
\bea
\rho_{\rm lead}(v) &=& \rho(v)\flead(v) \nonumber\\[2mm]
&=& \frac{\omega(v)}{\phi(v)}
\left\{ \,1\,-\,\int_{v_0}^v \!\dd v'
\left[\frac{\phi(v)}{\phi(v')}-1 \right] \omega(v')\tau(v') \right\}.
\label{exrholead}
\eea
By integrating (\ref{exrholead}) on $v$ one obtains
\beq
\brholead = \int_{v_0}^\infty \!\dd v \frac{\omega(v)}{\phi(v)}
\left\{ \,1\,-\,\omega\int_{v_0}^v \!\dd v'
\left[\frac{\phi(v)}{\phi(v')}-1 \right] P(v')\tau(v') \right\}.
\label{extrhototp}
\eeq
We obviously have the relation
\beq
\brholead = \brho-\brhofoll\,.
\label{sumrulerhotot}
\eeq
The quantities $\rholead(v)$ and $\brholead$ are also equal to the density
of platoons of velocity $v$ and the total platoon density, respectively. 

We may eliminate $\phi(v)$ from 
(\ref{exrholead}) by substituting for it
the solution (\ref{solchiv}).
The resulting expression can be simplified importantly.
In appendix \ref{secArholead} we show that after substantial rewriting
it takes the form
\beq
\rho_{\rm lead}(v) = \frac{\omega(v)}
{v\,+\,\int_{v_0}^{v}\!\dd v'\,(v-v')\omega(v')\tau(v')}\,,
\label{exrholeadsim}
\eeq
which relates $\rho_{\rm lead}(v)$ directly to $\omega(v)$, 
without the intervention of $\phi(v)$. 
We retain (\ref{exrholeadsim}) for later use. It is employed, moreover, in
appendix \ref{secappexactsol} to obtain the analytic expression for 
$\rho_{\rm lead}(v)$ in the case where $\omega(v)$ is a block distribution.
\vspace{5mm}

%%%%%%%%%%%%%%%%%%%%%%%%%%%%%%%%%%%%%%%%%%%%%%%%%%%%%%%%%%%%%%%%%%%%%%%%%%%%%%

{\it Platoon length distribution. --\,\,}
We consider a leading vehicle having a velocity that we now
call $v$ (it was called $v'$ above) and ask about the distribution of
the number $n$ of vehicles in its platoon.

At any instant of time,
in a spatial interval of length $L$ the total average number of
leading vehicles of natural velocity in $[v,v+\dd v]$ is 
$L\rholead(v)\dd v = L[\rho(v)-\rhofoll(v)]\dd v$. 
The average
total number of vehicles in this interval that are queuing behind leaders of
natural velocity in $[v,v+\dd v]$ is $L\sigmafoll(v)\dd v$.
Hence a leading vehicle of natural velocity $v$ is followed by an average
number 
of vehicles $\nu(v)$ given by the ratio of these two numbers,
\beq
\nu(v)=\frac{\sigmafoll(v)}{\rho(v)-\rhofoll(v)}
=\frac{\sigmafoll(v)}{\rholead(v)}\,.
\label{nuv}
\eeq
The number 
\beq
n(v)=1+\nu(v)
\label{defnv}
\eeq
is the average length of a platoon of velocity $v$.
We have the sum rules 
\bea
\int_{v_0}^\infty\!\dd v\,\rholead(v)\,\nu(v) &=& \brhofoll\,,
\nonumber\\[2mm]
\int_{v_0}^\infty\!\dd v\,\rholead(v)\,n(v) &=& \brho\,.
\label{sumrulepl}
\eea
The platoon length $\bar{n}$ averaged on platoons of arbitrary
velocities present in a given interval is equal to
\bea
\bar{n} &=& 1+\bar{\nu} 
\nonumber\\[2mm]
&=& 1\,+\,
\int_{v_0}^\infty \!\dd v\, \frac{\rho_{\rm lead}(v)\nu(v)}{\brholead}\,
\nonumber\\[2mm]
&=& \frac{\brho}{\brholead}\,,
\label{defnuav}
\eea
where we used (\ref{nuv}), (\ref{sigmarho}), and (\ref{sumrulerhotot}). 

A more detailed analysis of the composition of platoons
is possible along the same lines.
The total number of vehicles of natural velocity in $[v, v+\dd v]$
queuing behind leaders of natural velocity in $[v',v'+\dd v']$ is equal to
$L\rho(v,v')\dd v\dd v'$. Hence a leading vehicle of natural velocity
$v'$ is followed by an average number $\nu(v,v')\dd v$ of vehicles of natural
velocity in $[v, v+\dd v]$ given by
\beq
\nu(v,v')\dd v=\frac{\rho(v,v')\dd v}
{\rho(v')-\rhofoll(v')}\,.
\label{nuv1v}
\eeq
We have
\beq
\nu(v')=\int_v^\infty\!\dd v\,\nu(v,v').
\label{nuv1}
\eeq

Certain questions that one may ask require knowledge not only of the
the average platoon length (\ref{nuv1}), but of their probability distribution.
Let $p_n(v)$ be the probability that a randomly picked leading vehicle of 
velocity $v$ is heading an $n$-platoon $(n=1,2,\ldots)$.
The work done so far does not allow us to write down an expression for
$p_n(v)$. If however we make the additional assumption% 
%%%%%%%%%%
\footnote
{This is most likely an approximation.}
%%%%%%%%%%
that the followers are distributed randomly over the leaders,
then $p_n(v)$ is given by the Poisson distribution
\beq
p_n(v) = \ee^{-\nu(v)}\,\frac{[\nu(v)]^{n-1}}{(n-1)!}\,, \qquad n=1,2,\ldots
\label{pnv}
\eeq
This distribution is required, for example, when we ask about
`true' platoons, defined as platoons of $n\geq 2$
vehicles. Using (\ref{pnv}) we immediately find
that the density $\rho_{\rm trpl}(v)$ of true platoons of velocity $v$ 
is equal to
\bea
\rho_{\rm trpl}(v) &=& \rholead(v)[1-p_1(v)] \nonumber\\[2mm]
                   &=& \left[ \rho(v)-\rhofoll(v) \right] 
\left[1-\exp\left( -\frac{\sigmafoll(v)}{\rho(v)-\rhofoll(v)}\right)\right].
\label{}
\eea
We will not pursue these matters any further here.

%%%%%%%%%%%%%%%%%%%%%%%%%%%%%%%%%%%%%%%%%%%%%%%%%%%%%%%%%%%%%%%%%%%%%%%%%%%%%

\subsection{Circular traffic lane}
\label{seccircular}

For the purpose of simulation or for theoretical considerations
it may be more convenient to deal with periodic boundary conditions (PBC).
We consider in this section the changes that this brings about
with respect to the earlier case of open-ended boundary conditions (OBC).

We consider one-lane traffic on a circular lane of circumference $L$.
The spatial density $\rho(v)\dd v = \brho R(v)\dd v$ 
of vehicles with natural velocity in
$[v,v+\dd v]$ is now the prescribed control function.
We write again $w=\phi(v)$ for the effective velocity of a vehicle of natural 
velocity $w$. 
In this case the question is to find an equation that 
expresses $\phi(v)$ in terms of $\rho(v)$.
A shortcut way to do so is to assume that imposing boundary
conditions along a different axis
does not lead to a new type of stationary state and that the key
equation
\beq
\rho(v)=\frac{\omega(v)}{\phi(v)}\,,
\label{extrhovPBC}
\eeq
derived for OBC, remains valid for PBC with the same function $\phi(v)$.
When we use (\ref{extrhovPBC}) to eliminate $\omega(v)$ in favor of $\rho(v)$
from equation (\ref{eqnphiv0}) 
we obtain
\beq
\phi(v)=v-\int_{v_0}^v\!\dd v'\,
\rho(v')\tau(v')(v-v')\left[\phi(v)-\phi(v')\right],
\label{key4eq}
\eeq
where $\rho(v)$ is given and $\phi(v)$ is the function to be solved.
In appendix \ref{appsecphiPBC} we present a first-principle derivation 
of (\ref{key4eq}) that does not make use of this shortcut.
%%% Any pair $(\phi(v),\rho(v))$ that solves (\ref{key4eq})
%%% defines a pair $(\phi(v),\omega(v))$ that solves (\ref{eqnphiv0}), 
%%% and {\it vice versa}. 

In order to solve equation (\ref{key4eq}) we set
\bea
\Phi(v) &\equiv& v+\int_{v_0}^v\!\dd v'\,\rho(v')\tau(v')(v-v')\phi(v'),
\label{defPhiv}\\[2mm]
h(v) &\equiv& 1+\int_{v_0}^v\!\dd v'\,\rho(v')\tau(v')(v-v').
\label{defhv}
\eea
in which $\Phi(v)$ is unknown. Then\\[2mm]
\beq
\begin{array}{ll}
\Phi'(v)=1+\int_{v_0}^v\!\dd v'\,\rho(v')\tau(v')\phi(v'), \phantom{XXX}&
\Phi^{\prime\prime}(v) = \rho(v)\tau(v)\phi(v),\\[4mm]
h'(v) =\int_{v_0}^v\!\dd v'\,\rho(v')\tau(v'), &
h^{\prime\prime}(v) = \rho(v)\tau(v),
\end{array}\\[2mm]
\label{exPhihpv}
\eeq
which are such that $h(v_0)=1$ and $h'(v_0)=0$.
In terms of $\Phi$ and $h$ equation (\ref{key4eq}) takes the form
\beq
\frac{\Phi^{\prime\prime}(v)}{\Phi(v)} = \frac{h^{\prime\prime}(v)}{h(v)}\,.
\label{key5eq}
\eeq
The desired solution must satisfy the boundary conditions
$\Phi(v_0)=v_0$ and $\Phi'(v_0)=1$.
One obtains 
\beq
\Phi(v)=h(v)\left[ v_0 + \int_{v_0}^v\!\,\frac{\dd v'}{h^2(v')}\right],
\label{solPhiv}
\eeq
as may be verified by direct substitution of (\ref{solPhiv}) in (\ref{key5eq}).
Using (\ref{solPhiv}) and (\ref{exPhihpv}) we then arrive at the
desired expression for $\phi(v)$ in terms of $\rho(v)$,
\bea
\phi(v) &=& \frac{\Phi^{\prime\prime}(v)}{h^{\prime\prime}(v)} 
\nonumber\\[2mm]
&=& v_0 + \int_{v_0}^v\!\,\frac{\dd v'}{h^2(v')} \nonumber\\[2mm]
&=& v_0\,+\,\int_{v_0}^v\! \dd v'\,\left[\,
\,1\,+\,\int_{v_0}^{v'}\!\dd v^{\pp}\,\rho(v^{\pp})\tau(v^{\pp})(v'-v^{\pp}) 
\right]^{-2}.
\label{solphivpbc}
\eea
All relations found before involving $\rho_{\rm lead}$ and $\rho_{\rm foll}$
remain valid with $\omega(v)$ replaced by $\rho(v)\phi(v)$. 
In particular, one may rewrite (\ref{exrholead}) as
\beq
\rho_{\rm lead}(v) = \rho(v)\left[\,1\,-\,\int_{v_0}^v\!\dd v'\,
[\phi(v)-\phi(v')]\rho(v')\tau(v')\right].
\label{exrholeadpbc}
\eeq
By substituting (\ref{solphivpbc}) in (\ref{exrholeadpbc})
one may eliminate $\phi(v)$ and obtains after simplifying 
\beq
\rho_{\rm lead}(v) = \frac{\rho(v)}
{1\,+\,\int_{v_0}^v\!\dd v'\,\rho(v')\tau(v')(v-v')}\,.
\label{exrholeadsimpbc}
\eeq
This is the PBC analog of the OBC equation (\ref{exrholeadsim}).

%%%%%%%%%%%%%%%%%%%%%%%%%%%%%%%%%%%%%%%%%%%%%%%%%%%%%%%%%%%%%%%%%%%%%%%%%%%%%

\subsection{The moments $X$ and $Y$ of $\rho_{\rm lead}(v)$}
\label{secmoments}

The following notation will be useful later on.
We will let $X$ and $Y$ denote the zeroth and first moment,
respectively, of $\rholead(v)$, that is,
\begin{subequations}\label{momentsrholead}
\beq \label{momentsrholeadv0}
   X
= \int_{v_0}^\infty\!\dd v\, \rholead(v)
=\brholead\,,
\eeq
\beq \label{momentsrholeadv1}
   Y
= \int_{v_0}^\infty\!\dd v\, v\rholead(v)
=\brholead\la v\ra_{\rm lead}\,,
\eeq
\end{subequations}
so that $\la v\ra_{\rm lead}=Y/X$ is the average platoon velocity.

%%%%%%%%%%%%%%%%%%%%%%%%%%%%%%%%%%%%%%%%%%%%%%%%%%%%%%%%%%%%%%%%%%%%%%%%%%%%%

\subsection{Upper density limit}
\label{secvalidity}

The model of this section obviously contains simplifications that may
all be a matter of discussion. One of these, however, needs to be
addressed in view of our ultimate conclusions.
The fact that in this model
the vehicles and platoons are all considered to have zero length,
sets a limit on its validity when the total vehicle density increases.
We may formulate the limit of validity as the condition that the
typical platoon length $\ell_{\rm pl}$ be much smaller than the
typical distance $d_{\rm interpl}$
between two successive platoons. 
Now we have $d_{\rm interpl}=1/\rho_{\rm lead}$, and may write
$\ell_{\rm pl}=\bar{n}d_{\rm interveh}$, where $d_{\rm interveh}$ is
the typical distance between two successive vehicles in a platoon
(and includes the typical length of a vehicle).
Hence the condition $\ell_{\rm pl} \ll d_{\rm interpl}$ becomes
\beq
 \bar{n}\,d_{\rm interveh} \ll \frac{1}{\brho_{\rm lead}}.
\label{validity1}
\eeq
When with the aid of (\ref{defnuav}) we eliminate $\bar{n}$ from 
(\ref{validity1}) we obtain the limit of validity
\beq
\brho\,d_{\rm interveh} \ll 1.
\label{validity2}
\eeq
Here $d_{\rm interveh}$ is a parameter that is external to the present
model (where it has in fact been set equal to zero), but may be estimated
from observations of real-life traffic. We will check
the status of this condition numerically in section \ref{secvaliditynum}.

%%%%%%%%%%%%%%%%%%%%%%%%%%%%%%%%%%%%%%%%%%%%%%%%%%%%%%%%%%%%%%%%%%%%%%%%%%%%%
%%%%%%%%%%%%%%%%%%%%%%%%%%%%%%%%%%%%%%%%%%%%%%%%%%%%%%%%%%%%%%%%%%%%%%%%%%%%%

\section{A model for two opposite lanes}
\label{secopposite}

%%%%%%%%%%%%%%%%%%%%%%%%%%%%%%%%%%%%%%%%%%%%%%%%%%%%%%%%%%%%%%%%%%%%%%%%%%%%%

We now consider two lanes $A$ and $B$
with opposite traffic flows.
In the general case of nonidentical traffic conditions on the two lanes
there will be two control functions,
$\omega_A(v)=\bomega_AP_A(v)$ and $\omega_B(v)=\bomega_BP_B(v)$,
and two queuing time functions, $\tau_A(v)$ and $\tau_B(v)$.
In the one-lane problem of the preceding section the queuing time
function was considered given.
In this section we will aim at expressing
the queuing time $\tau_A(v)$ experienced by the
traffic in lane $A$ in terms of the traffic
conditions in the opposite lane $B$, and {\it vice versa}.
First, in subsection \ref{secgamma}, we introduce the concept of the
``encounter rate'' between a vehicle in one lane and the platoons
coming in the opposite direction. This builds straightforwardly
on the work of section \ref{secsinglelane}. 
Secondly, in subsection \ref{sectau}, we need a mean-field type
hypothesis: we assume that when attempting an overtaking maneuver,
{\it each vehicle in one lane is subject to the
average effect of all vehicles in the opposite lane}.
This will result in a set of equations 
in which the two lanes appear coupled. The structure of these
equations is briefly discussed in section \ref{seccoupled}.
From these coupled equations we will then in section \ref{secnumerical}
be able to solve self-consistently
-- albeit only numerically --
the functions $\tau_A(v)$ and $\tau_B(v)$.

%%%%%%%%%%%%%%%%%%%%%%%%%%%%%%%%%%%%%%%%%%%%%%%%%%%%%%%%%%%%%%%%%%%%%%%%%%%%%

\subsection{The encounter rates  $\gamma_A$ and $\gamma_B$}
\label{secgamma}

As a preliminary let us consider two independent single-lane models
$A$ and $B$,
one for each of the two directions.
Let $\gamma_{A}(v)$ [or $\gamma_B(v)$]
denote the rate at which a platoon of velocity $v$ in
lane $A$ [or in lane $B$] 
encounters platoons in the opposite direction\footnote% 
{We take all velocities positive, so that two vehicles of velocities $v_a$
and $v_b$ in opposite directions have a relative velocity $v_a+v_b$.};
$\gamma_{A}(v)$ and $\gamma_B(v)$ are necessarily an increasing function of $v$.
The same methods as used in the single-lane model
lead straightforwardly
to expressions for these encounter rates, as we will see now.

During a time interval $[0,T]$
a marked platoon in lane $A$ initially supposed in the origin
will encounter all lane $B$ platoons 
of velocity $v_b$
that are initially in the space interval $[0,(v+v_b)T]$.
Let $N_{B,\rm lead}(v,v_b)\dd v_b$ be the average number of platoons
on lane $B$  
encountered by the marked one on lane $A$ that have their velocity in
$[v_b,v_b+\dd v_b]$. 
Then 
\beq
N_{\rm lead}(v,v_b)\dd v_b = (v+v_b)T \times \rhoBlead(v_b)\dd v_b\,.
\label{eqa}
\eeq
The total number $\gamma_A(v)T$ of platoons encountered by the marked
platoon is
\beq
\gamma_A(v)T = \int_{v_0}^\infty \!\dd v_b\,N_{B,\rm lead}(v,v_b)
\label{eqb}
\eeq
and therefore from (\ref{eqa}) and (\ref{eqb})
\beq
\gamma_A(v) = \!\int_{v_0}^\infty \!\dd v_b\,(v+v_b)\,\rhoBlead(v_b).
\label{gammav0}
\eeq
Hence, in the notation of equation (\ref{momentsrholead}) augmented with the
appropriate lane indices,
\begin{subequations}\label{gammav}
\beq\label{gammaAv}
\gamma_A(v) = \brhoBlead \left[ v+\langle v\raBlead \right]
            =X_Bv + Y_B,
\eeq
and, by a permutation of the indices,
\beq\label{gammaBv}
\gamma_B(v) = \brhoAlead \left[ v+\langle v\raAlead \right]
            = X_Av + Y_A.
\eeq
\end{subequations}
We conclude this subsection by a remark.

{\it Remark. --\,\,} Let $\bgamma_{A,B}$ be defined as the averages
of the $\gamma_{A,B}(v)$ of equations (\ref{gammav}) 
with respect to all platoons in lanes $A$ and $B$, respectively.
Explicitly,
\begin{subequations}\label{gammaba}
\beq\label{gammaAba}
\bgamma_{A} = \brhoBlead
[\la v\raAlead + \la v\raBlead] 
= X_BY_A/X_A + Y_B,
\eeq
\beq\label{gammaBba}
\bgamma_{B} = \brhoAlead
[\la v\raAlead + \la v\raBlead]
= X_AY_B/X_A + Y_A.
\eeq
\end{subequations}
Then the expression
\bea
\brhoAlead\bgamma_A 
=\brhoBlead\bgamma_B 
&=&\brhoAlead\brhoBlead
[\la v\raAlead + \la v\raBlead] \nonumber\\[2mm]
&=& X_AY_B + X_BY_A\,,
\label{encounters}
\eea
which is symmetric under exchange of the two lanes, 
represents the number of encounters per unit of time and
unit of road length of platoons traveling in opposite directions.

%%%%%%%%%%%%%%%%%%%%%%%%%%%%%%%%%%%%%%%%%%%%%%%%%%%%%%%%%%%%%%%%%%%%%%%%%%%%%

\subsection{Coupling the two lanes}
\label{sectau}

In order to establish the equations that
couple the two lanes, we reason as follows. Let 
$\tau_0$ be the elementary time interval
spent by an overtaking vehicle in the lane that is not its own.
We consider $\tau_0$ as a fixed constant, in practice of the order of 
15 seconds. We now consider that
a vehicle queuing behind another one will overtake as soon as
the platoons arriving in the opposite direction allow for a time
interval at least equal to $\tau_0$.
The times $\tau_A(v)$ and $\tau_B(v)$ are the queuing times that
result from this condition.
Hence they are determined by $\tau_0$
{\it and\,} the encounter rates with the traffic in the opposite lane.
We will now construct explicit expressions for $\tau_{A,B}(v)$.

In order to relate $\gamma_{A,B}(v)$ to $\tau_{A,B}(v)$ 
we introduce the additional hypothesis
(briefly commented upon in section \ref{secmathematical})
that the platoons arrive at random times.
The result, obtained in appendix \ref{secapptau}, then is that there is a
function $F(x)$, increasing monotonously from $0$ to $\infty$
on the positive real axis, such that 
\beq
\tau_{A,B}(v)=\tau_0 F\big (\gamma_{A,B}(v)\tau_0 \big), 
\label{restau}
\eeq
where we recall that the $\gamma_{A,B}$ are given by equation (\ref{gammav}).
Appendix \ref{secapptau} yields the explicit expression
$F(x)=(\ee^x-1-x)/x$, but
in what follows
no doubt any other $F(x)$ with qualitatively the same behavior would
lead to qualitatively the same results.

Equation (\ref{restau}) is the desired relation that
expresses the queuing times in one lane as a function of the
rate of encounters with platoons in the opposite lane. 
It represents a coupling of the mean-field type: the traffic in one
lane reacts to the {\it average\,} traffic conditions in the other lane.

%%%%%%%%%%%%%%%%%%%%%%%%%%%%%%%%%%%%%%%%%%%%%%%%%%%%%%%%%%%%%%%%%%%%%%%%%%%%%

\subsection{The two-lane equations}
\label{seccoupled}

Let the lane $B$
averages $X_B=\brhoBlead$ and $Y_B=\brhoBlead\langle v\raBlead$ be known.
Because of (\ref{gammaAv}) and (\ref{restau}) we then
know $\tau_A(v)$ in terms of $X_B$ and $Y_B$\,.
We next use this expression for $\tau_A(v)$ in the single-lane
theory of section \ref{secsinglelane}. 
When substituted in equation (\ref{exrholeadsim}) 
[in the case of OBC] or in
equation (\ref{exrholeadsimpbc}) [in the case of PBC]
it yields $\rhoAlead(v)$, from which follow by (\ref{momentsrholead}) the
lane $A$ averages $X_A=\brhoAlead$ and $Y_A=\brhoAlead\langle v\raAlead$,
again in terms of $X_B$ and $Y_B$.
We may write concisely 
\begin{subequations}\label{stat}
\beq\label{statA}
X_A=f_A(X_B,Y_B),  \qquad Y_A=g_A(X_B,Y_B).
\eeq
A permutation of indices yields
\beq\label{statB}
X_B=f_B(X_A,Y_A),  \qquad Y_B=g_B(X_A,Y_A).
\eeq
\end{subequations}
This system of equations may be solved numerically,
which will be done for a few examples in section \ref{secnumerical}.

The indices on $f$ and $g$ 
are necessary because these functions depend on $\omega_A$ and
$\omega_B$, respectively, in the case of OBC, and on $\rho_A$ and $\rho_B$,
respectively, in the case of PBC.
For equal control functions, $\omega_A(v)=\omega_B(v)$
or $\rho_A(v)=\rho_B(v)$ (which we will call the `symmetric' case) 
we have $f_A=f_B\equiv f$ and $g_A=g_B\equiv g$ and hence
\begin{subequations}\label{statsymm}
\beq\label{statAsymm}
X_A=f(X_B,Y_B),  \qquad Y_A=g(X_B,Y_B),
\eeq
\beq\label{statBsymm}
X_B=f(X_A,Y_A),  \qquad Y_B=g(X_A,Y_A).
\eeq
\end{subequations}
The symmetric case will be the one of main interest in what follows.
If, in the symmetric case, we wish to just obtain a symmetric solution 
$X_A=X_B=X$ and $Y_A=Y_B=Y$ of (\ref{statsymm}), 
we can simplify that equation further and get $X=f(X,Y)$ and $Y=g(X,Y)$.
However, under certain conditions the symmetric case
will be seen to allow for solutions with spontaneously broken
symmetry; to find these the full equations (\ref{statsymm}) are necessary.

%%%%%%%%%%%%%%%%%%%%%%%%%%%%%%%%%%%%%%%%%%%%%%%%%%%%%%%%%%%%%%%%%%%%%%%%%%%%%

\section{Numerical solution of the two-lane equations}
\label{secnumerical}

The two-lane equations may be solved numerically.
The procedure runs as follows.
We assume suitably chosen initial values for the pair
$(X_B,Y_B)$ and then alternatingly update $(X_A,Y_A)$ and $(X_B,Y_B)$
by applying equations (\ref{statA}) and (\ref{statB}).
In practice equations (\ref{exrholeadsim}) and (\ref{exrholeadsimpbc}), 
which are at the heart of the iteration procedure, have to be discretized.
We use the natural discretization, that is, the form that these
equations take when the control functions $\omega_{A,B}(v)$ or
$\rho_{A,B}(v)$ are sums of Dirac delta's. This discretized version
is described in appendix \ref{secappdiscr}, where 
equation (\ref{exrholeadsim}) has as its equivalent
in (\ref{exrholeadsimdiscr}) together with (\ref{defgdis}), and
(\ref{exrholeadsimpbc}) has as its equivalent in
(\ref{exrholeaddissimpbc}) together with (\ref{defhdis}).

In all cases that we have considered 
there is convergence to a solution,
which appears either as a fixed point with $(X_A,Y_A)=(X_B,Y_B)$ or as a fixed
two-cycle alternating between $(X_A,Y_A)$ and $(X_B,Y_B)$.
The convergence slows down near the critical point (see below),
but we have not found it worthwhile to improve the algorithm.

%%%%%%%%%%%%%%%%%%%%%%%%%%%%%%%%%%%%%%%%%%%%%%%%%%%%%%%%%%%%%%%%%%%%%%%%%%%%%

\subsection{Symmetric lanes with OBC}
\label{secexamples}

We consider first the symmetric case with identical {\it a priori\,}
traffic conditions for the two lanes, that is, equal distributions
$\omega_A(v)=\omega_B(v)$, which we may therefore
denote by $\omega(v)=\bomega P(v)$. We recall that $\bomega$ is the
total intensity of the traffic flow, that is,
the rate of entrance of vehicles, irrespective of their velocity $v$,
at the point $x=0$ of the lane section under consideration.
For the distribution $\omega(v)$ we take a truncated Gaussian 
discretized on $N+1$ points
$v_i=v_0+i\Delta v$ with  $i=0,1,\ldots,N$. 
Explicitly,
\begin{subequations}\label{example}
\beq\label{exampleomegav}
\omega(v) = \sum_{i=0}^{N}\omega_i\delta(v-v_i)
\eeq
with $\omega_i \equiv \bomega p_i$ and 
\beq\label{truncGauss}
p_i\,=\,{\cal N}\exp\left(-\frac{(v_i-v_{\rm c})^2}{2\sigma^2}\right), 
\qquad i=0,1,\ldots,N,
\eeq
where $v_{\rm c}$ is the peak velocity
and ${\cal N}$ the normalization constant required to satisfy 
$\sum_{i=0}^Np_i=1$.
We have studied numerically the example with
\beq\label{examplenum}
\begin{array}{lll}
v_0=60\,\mbox{km/h},                     & \sigma=10\,\mbox{km/h},
\phantom{XX}                             & \Delta v=1\,\mbox{km/h},\\[2mm]
v_{\rm c}=90\,\mbox{km/h},\phantom{XXX}  & N=60,
                                         &
\end{array}
\eeq
\end{subequations}
so that the largest occurring velocity is $v_{\rm m}=120$\,km/h.
When averaged over all vehicles in a road section,
and in the absence of interaction between the lanes,
equation (\ref{examplenum}) yields a mean vehicle velocity of 
$\la v\ra^{(0)}=88.9$\,km/h (it is less than the average velocity of
$90$\,km/h since slower vehicles stay longer
in the section than faster ones). 

\begin{figure}
\begin{center}
\scalebox{.55}
%{\includegraphics{XY107i.eps}}
{\includegraphics{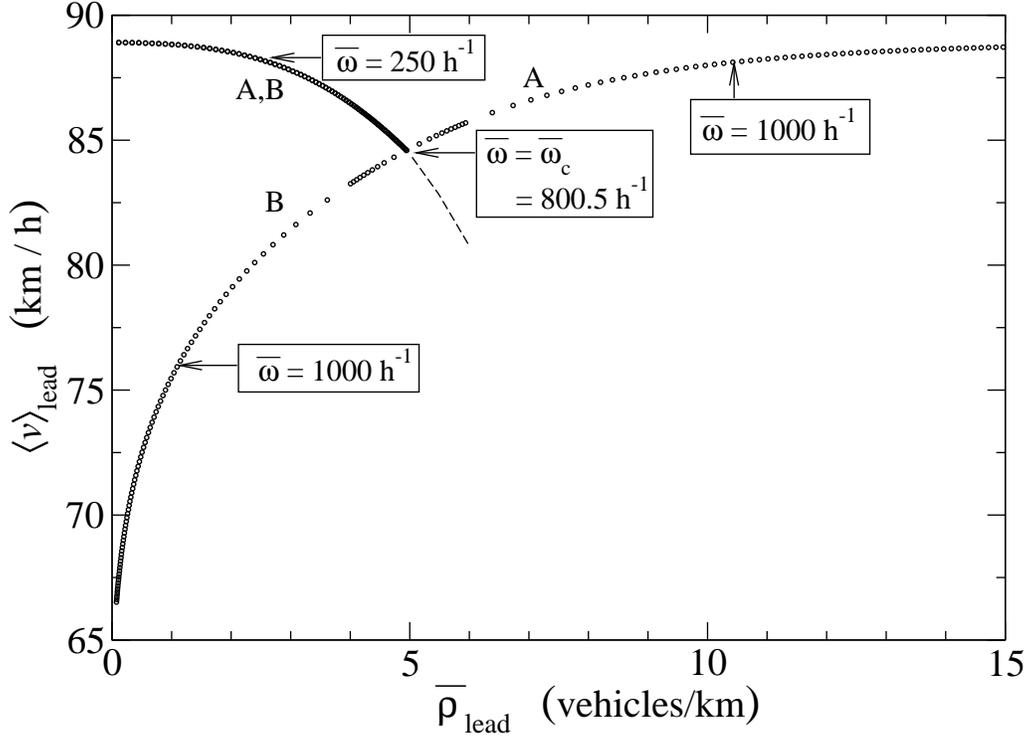}}
\end{center}
\caption{\small 
(OBC)
For the velocity distribution $\omega(v)=\bomega P(v)$ 
given by equations (\ref{example}),
as the control parameter $\bomega$ is varied,
the two-lane equations (\ref{statsymm}) yield a
locus of solutions of solutions has been plotted here 
in the plane of coordinates
$(X,YX^{-1})=(\brho_{\rm lead},\la v\ra_{\rm lead})$.
The control parameter
increases along the branch marked `A,B' from $\bomega=0$ 
in the upper left corner to 
around $\bomega=\bomega_{\rm c}\approx 800$ h$^{-1}$ 
at the bifurcation point, and, going outward from this
point, from $\bomega = \bomega_{\rm c}$ to around $\bomega=1400$ h$^{-1}$ along
each of the two branches marked A and B.
The dashed line represents schematically a branch of unstable solutions.}
\label{symXY}
\end{figure}

For this example we determined 
the solutions $(X_A,Y_A)$ and $(X_B,Y_B)$ with the control parameter $\bomega$
varying from $\bomega=0$ to about $\bomega=1500$ vehicles/hour.
Figure \ref{symXY} shows the results represented in
the plane of coordinates $(X,YX^{-1})=(\brholead,\la v\ra_{\rm lead})$.
In the branch marked `A,B' the parameter $\bomega$ 
runs from zero in the upper lefthand corner to a critical value
$\bomega=\bomega_{\rm c}=800.5\,$h$^{-1}$ at a point where the curve
bifurcates.
The interpretation is that
for $\bomega<\bomega_{\rm c}$ the two lanes $A$ and $B$ have identical 
behavior; but
when $\bomega$ reaches its critical value,
the symmetric solution becomes unstable and
the symmetry between the lanes $A$ and $B$ is {\it spontaneously broken}.
From the bifurcation point two branches marked A and B 
emanate; in the figure $\bomega$ runs from $\bomega_{\rm c}$ to a
value of the order of $1500$ vehicles/hour along these branches.
Lane $B$ has a high total density of vehicles, many of which are
included in platoons, and the other lane $A$ has a lower density of
vehicles, the majority of which
are leading, and thus many of them must constitute 1-platoons.
The continuation of the `A,B' branch, where the symmetric solution is
unstable, has been indicated as a dashed line. 

\begin{figure}[h]
\begin{center}
\scalebox{.55}
%{\includegraphics{XY115b.eps}}
{\includegraphics{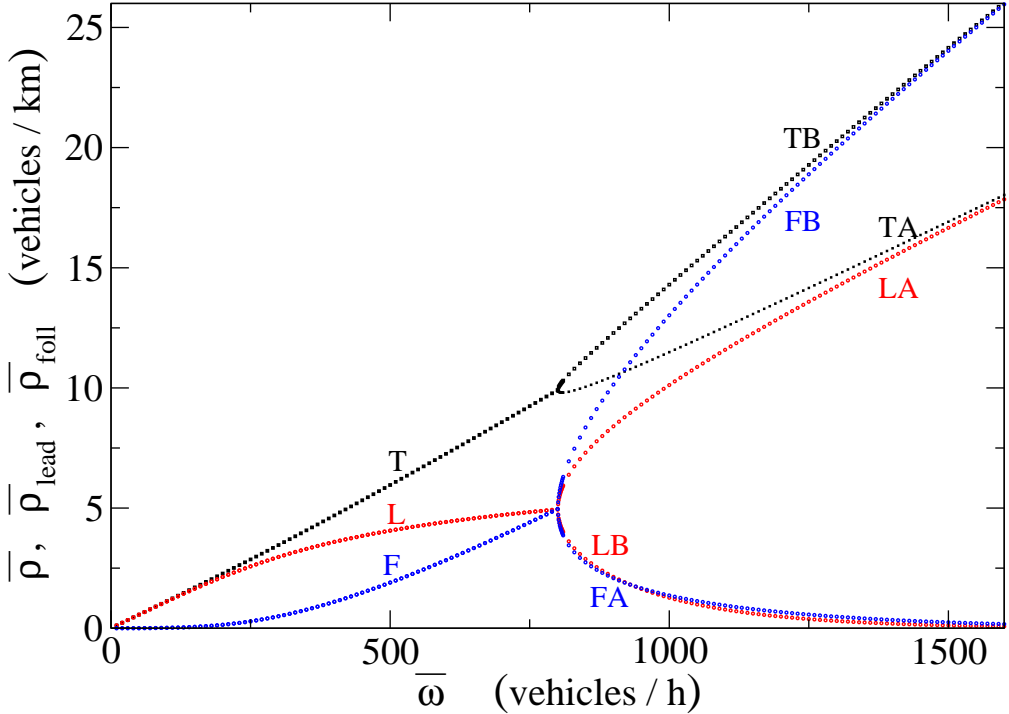}}
\end{center}
\caption{\small 
(OBC)
The total vehicle density (marked T),
  which is the sum of the leading vehicle density (L) and the 
  following vehicle density (F), plotted as a function of the
  flux $\bomega$
for the same distribution $\omega(v)$ (common to both lanes)
as in figure \ref{symXY}.
  At the critical flux
  $\bomega_{\rm c}=800.5$ vehicles/hour each of these three curves
  bifurcates into an $A$ and a $B$ branch, signaling spontaneous symmetry
  breaking between the two lanes.} 
\label{symdens}
\end{figure}

Figure \ref{symdens} shows, as a function of the control parameter
$\bomega$, the densities $\brho_{\rm lead}$ and $\brho_{\rm foll}$ 
of the leading and following vehicles, as well as their sum $\brho$.
Figures \ref{symn} and \ref{symv} show the corresponding graphs of
the average platoon length $\bar{n}$ and the average platoon 
velocity $\la v\ra_{\rm lead}$.
For small $\bomega$ we are in the weakly interacting
regime. The three densities shown in figure \ref{symdens} are linear in 
$\bomega$, with deviations from linearity becoming appreciable
when $\bomega$ attains values of a few hundred vehicles/hour.
We recall now that $\brho_{\rm lead}$ is also the density of platoons. 
Figure \ref{symdens} shows that
for small $\bomega$ the platoon density $\brho_{\rm lead}$ is
practically equal to the total vehicles density $\brho$, which means that
most vehicles constitute $1$-platoons by themselves.  
This is confirmed by the initial behavior of $\bar{n}$ in figure
\ref{symn}. Consistently, $\la v\ra_{\rm lead}$ in figure \ref{symv}
stays initially equal to its noninteracting value $\la v\ra^{(0)}$.

Beyond the small $\bomega$ regime
the average platoon velocity begins to decrease
due to faster vehicles getting trapped in slow platoons.
The instability occurs, in this example, when the average
platoon length has increased to about two cars per platoon and the
average platoon speed has gone down to about $85$\,km/h.
The interpretation of this symmetry breaking
is most easily derived from the large $\bomega$
behavior of the two branches.
As $\bomega$ increases to values well above
$\bomega_{\rm c}$, there occurs formation of a few very long platoons
behind vehicles having velocities close to the minimum velocity.
This explains why on branch $B$ asymptotically $\brho_{\rm lead}\to 0$
and $\la v\ra_{\rm lead}\to v_0$.
The small number of platoons in this lane have large spaces 
between them 
which allow the vehicles in lane $A$ to overtake almost without having
to queue. Hence asymptotically  on the $A$ branch
$\la v\ra_{\rm lead}\to\la v\ra^{(0)}$.
This interpretation is again confirmed by the figures
\ref{symdens}, \ref{symn}, and \ref{symv}.
Particularly striking is the rapid increase of the average platoon
length in lane $B$ when $\bomega$ increases in the supercritical regime.
It returns to the average vehicle velocity in the absence of interaction; 
in lane $B$
it tends asymptotically to the velocity $v_0 = 60$ km\,h$^{-1}$ 
of the slowest vehicles.

\begin{figure}
\begin{center}
\scalebox{.55}
%{\includegraphics{XY123h.eps}}
{\includegraphics{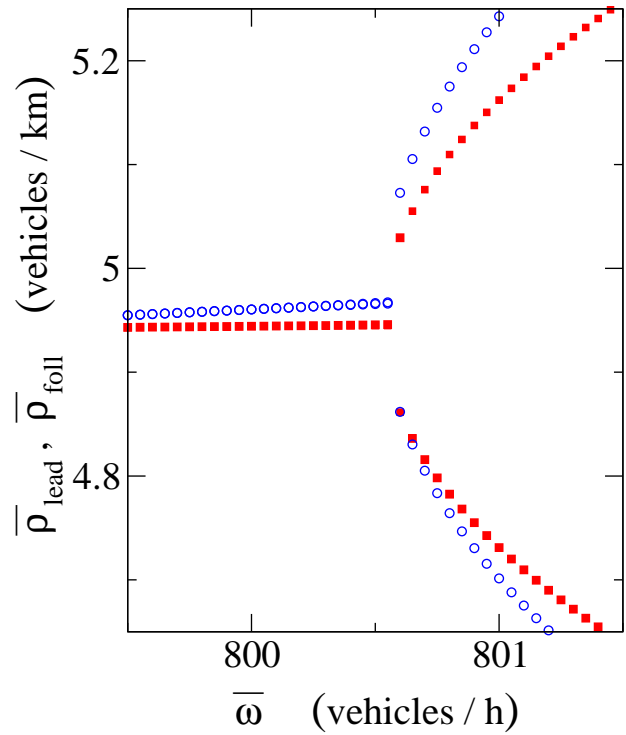}}
\end{center}
\caption{\small 
%symdensdet XY123h 
Detail of figure \ref{symdens}.
Closed squares: $\brho_{\rm lead}$. Open circles: $\brho_{\rm foll}$.}
\label{symdensdet}
\end{figure}

\begin{figure}
\begin{center}
\scalebox{.55}
%{\includegraphics{XY110e.eps}}
{\includegraphics{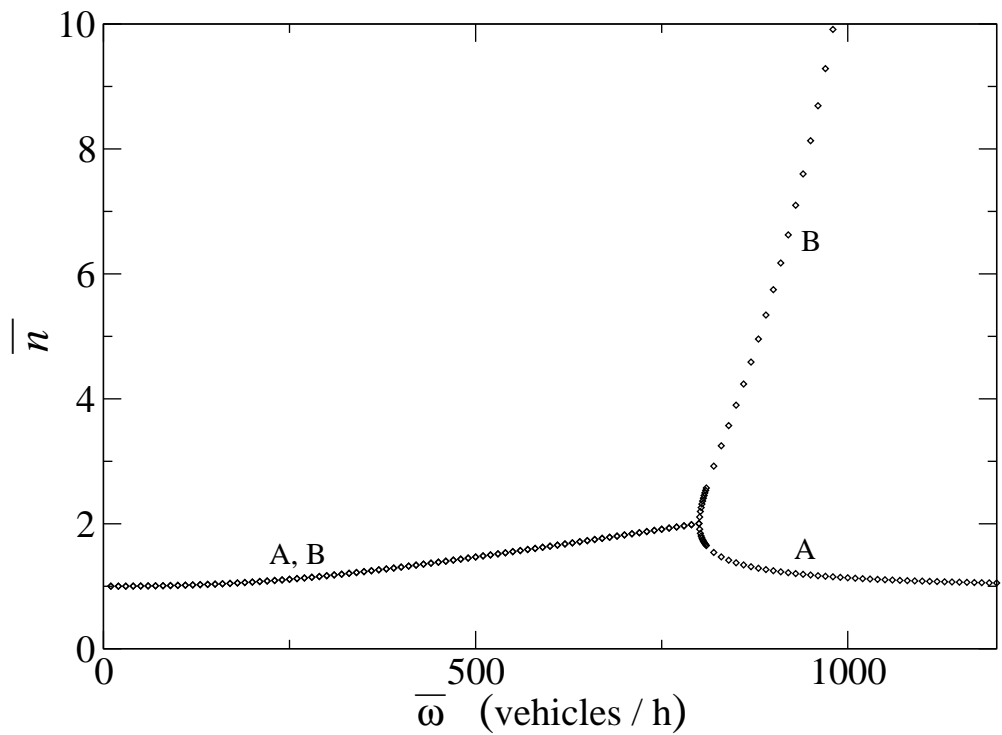}}
\end{center}
\caption{\small 
(OBC)
Average platoon length as a function of
  the flux $\bomega$
for the same distribution $\omega(v)$ (common to both lanes)
as in figure \ref{symXY}.
For $\bomega>\bomega_{\rm c}$ the platoon
  length in lane $B$ is seen to increase very rapidly with the flux.}
\label{symn}
\end{figure}

\begin{figure}
\begin{center}
\scalebox{.55}
%{\includegraphics{XY109e.eps}}
{\includegraphics{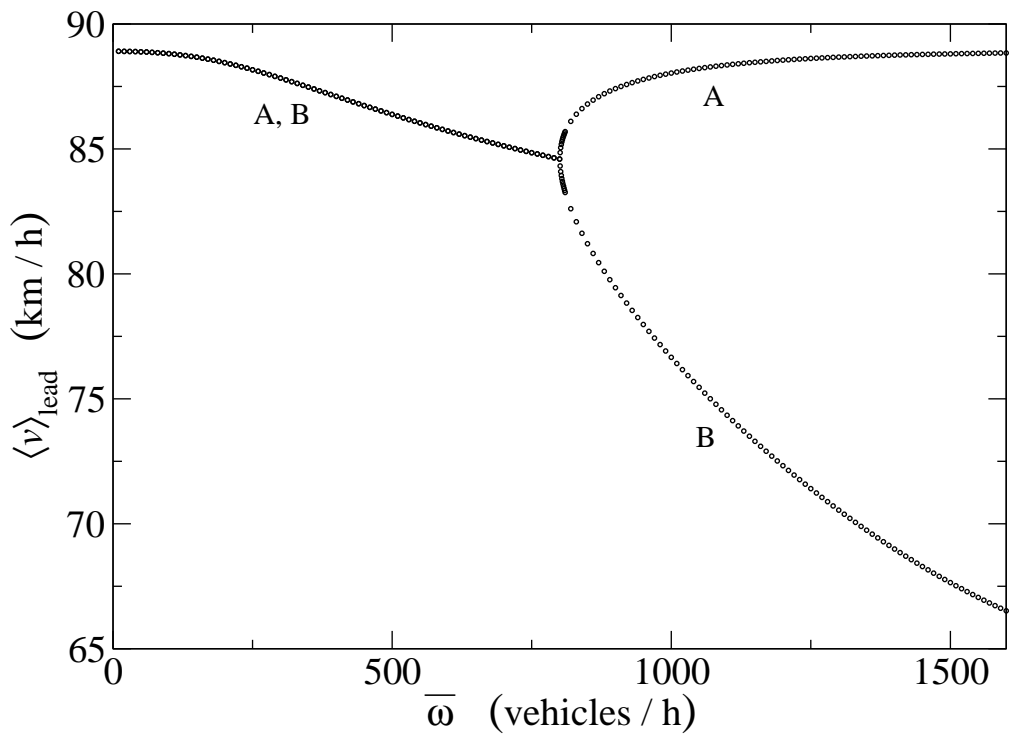}}
\end{center}
\caption{\small 
(OBC)
The average platoon velocity as a function of
  the flux $\bomega$ for the same distribution $\omega(v)$
(common to both lanes) as in figure \ref{symXY}.
For $\bomega>\bomega_{\rm c}$ the platoon
  velocity in lane $B$ is seen to decrease rapidly with the flux.}
\label{symv}
\end{figure}

We note that
the bifurcation points 
of $\brho_{\rm lead}$ and $\brho_{\rm foll}$ seem to coincide in
figure \ref{symdens}, but figure \ref{symdensdet} shows that they are
actually distinct.

\begin{figure}
\begin{center}
\scalebox{.55}
%{\includegraphics{XY114a.eps}}
{\includegraphics{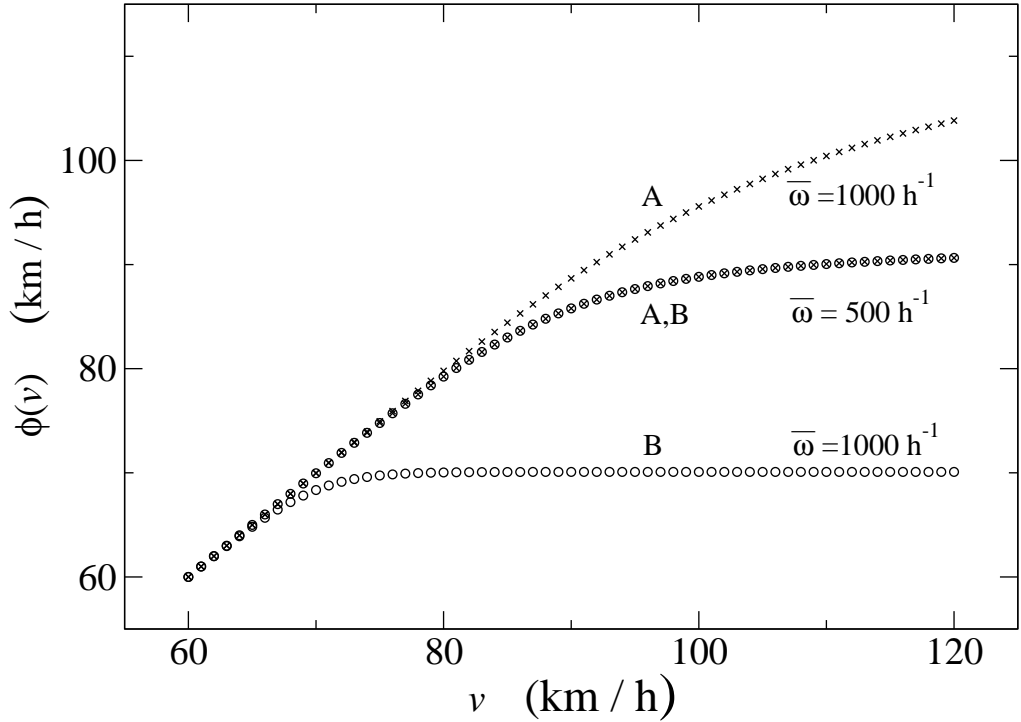}}
\end{center}
\caption{\small 
(OBC)
The effective velocity $\phi(v)$ of a vehicle
  as a function of its natural velocity $v$. The distribution
  $\omega(v)$ is the same as in figure \ref{symXY} and common to both
  lanes. At 
  the subcritical flux of $\bomega=500$ vehicles/hour 
  lanes $A$ and $B$ are described by a common
  curve, whereas at the supercritical flux $\bomega=1000$ vehicles/hour 
  they are described by two separate curves.}
\label{symphi}
\end{figure}

Finally in figure \ref{symphi}
we show the behavior of the effective velocity $w=\phi(v)$ for two
values of $\bomega$. 
They all have the expected behavior.
For the subcritical flux $\bomega = 500$ vehicles/hour
the functions $\phi_A(v)$ and $\phi_B(v)$ coincide.
For the supercritical value $\bomega = 1000$ vehicles/hour
they are distinct; the lane $A$ curve is not far from the straight line
$\phi(v)=v$ valid for the noninteracting case, and in the lane $B$ curve
the effective velocities are reduced to values not far from the
minimum velocity $v_0=60$\,km/h.

We have varied the distribution $P(v)$ and found that this does not
affect the qualitative features of the results that we discussed.

%%%%%%%%%%%%%%%%%%%%%%%%%%%%%%%%
%%%%%%%%%%%%%%%%%%%%%%%%%%%%%%%%
%%%%%%%%%%%%%%%%%%%%%%%%%%%%%%%%
%%%%%%%%%%%%%%%%%%%%%%%%%%%%%%%%

%%%%%%%%%%%%%%%%%%%%%%%%%%%%%%%%
%%%%%%%%%%%%%%%%%%%%%%%%%%%%%%%%
%%%%%%%%%%%%%%%%%%%%%%%%%%%%%%%%
%%%%%%%%%%%%%%%%%%%%%%%%%%%%%%%%

%%%%%%%%%%%%%%%%%%%%%%%%%%%%%%%%
%%%%%%%%%%%%%%%%%%%%%%%%%%%%%%%%
%%%%%%%%%%%%%%%%%%%%%%%%%%%%%%%%
%%%%%%%%%%%%%%%%%%%%%%%%%%%%%%%%

%%%%%%%%%%%%%%%%%%%%%%%%%%%%%%%%
%%%%%%%%%%%%%%%%%%%%%%%%%%%%%%%%
%%%%%%%%%%%%%%%%%%%%%%%%%%%%%%%%
%%%%%%%%%%%%%%%%%%%%%%%%%%%%%%%%

%%%%%%%%%%%%%%%%%%%%%%%%%%%%%%%%
%%%%%%%%%%%%%%%%%%%%%%%%%%%%%%%%
%%%%%%%%%%%%%%%%%%%%%%%%%%%%%%%%
%%%%%%%%%%%%%%%%%%%%%%%%%%%%%%%%

%%%%%%%%%%%%%%%%%%%%%%%%%%%%%%%%
%%%%%%%%%%%%%%%%%%%%%%%%%%%%%%%%
%%%%%%%%%%%%%%%%%%%%%%%%%%%%%%%%
%%%%%%%%%%%%%%%%%%%%%%%%%%%%%%%%

%%%%%%%%%%%%%%%%%%%%%%%%%%%%%%%%
%%%%%%%%%%%%%%%%%%%%%%%%%%%%%%%%
\begin{figure}
\begin{center}
\scalebox{.55}
%{\includegraphics{XY131c.eps}}
{\includegraphics{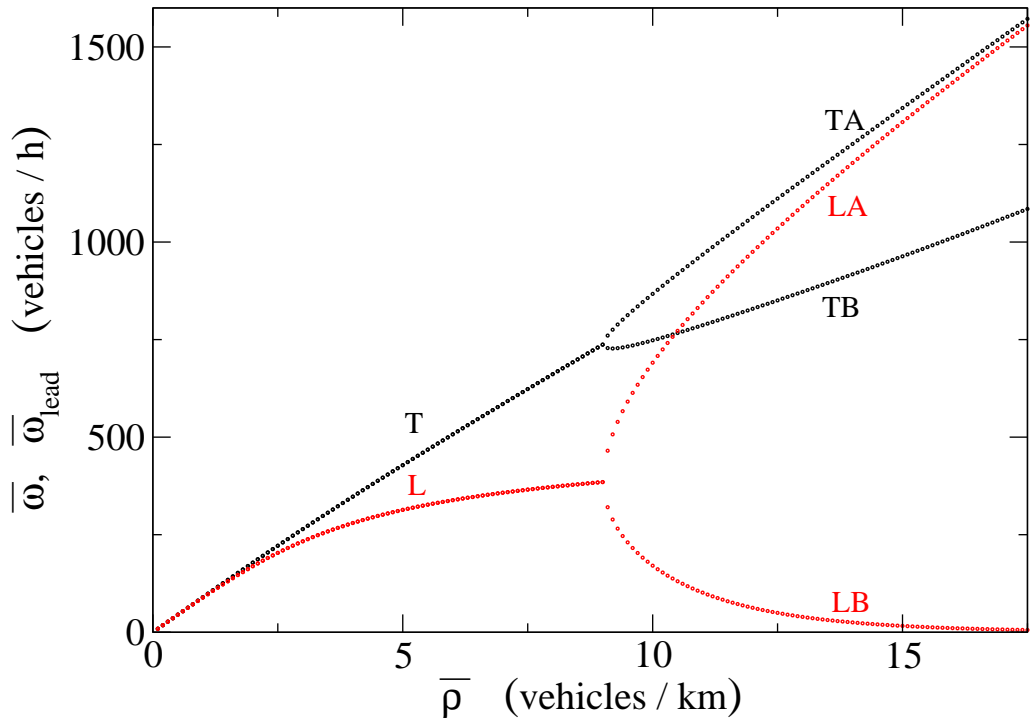}}
\end{center}
\caption{\small 
Case of periodic boundary conditions
(PBC) with vehicle density $\rho(v)=\brho R(v)$ and $R(v)$ a truncated
Gaussian described in the text.
Shown are the total vehicle flux $\bomega$ (marked T) and the
leading vehicle flux $\bomega_{\rm lead}$ (marked L) as a
function of the vehicle density $\brho$, which is here the control parameter.}
\label{symrhodens}
\end{figure}
%%%%%%%%%%%%%%%%%%%%%%%%%%%%%%%%
%%%%%%%%%%%%%%%%%%%%%%%%%%%%%%%%

%%%%%%%%%%%%%%%%%%%%%%%%%%%%%%%%
%%%%%%%%%%%%%%%%%%%%%%%%%%%%%%%%
\begin{figure}
\begin{center}
\scalebox{.55}
%{\includegraphics{XY116d.eps}}
{\includegraphics{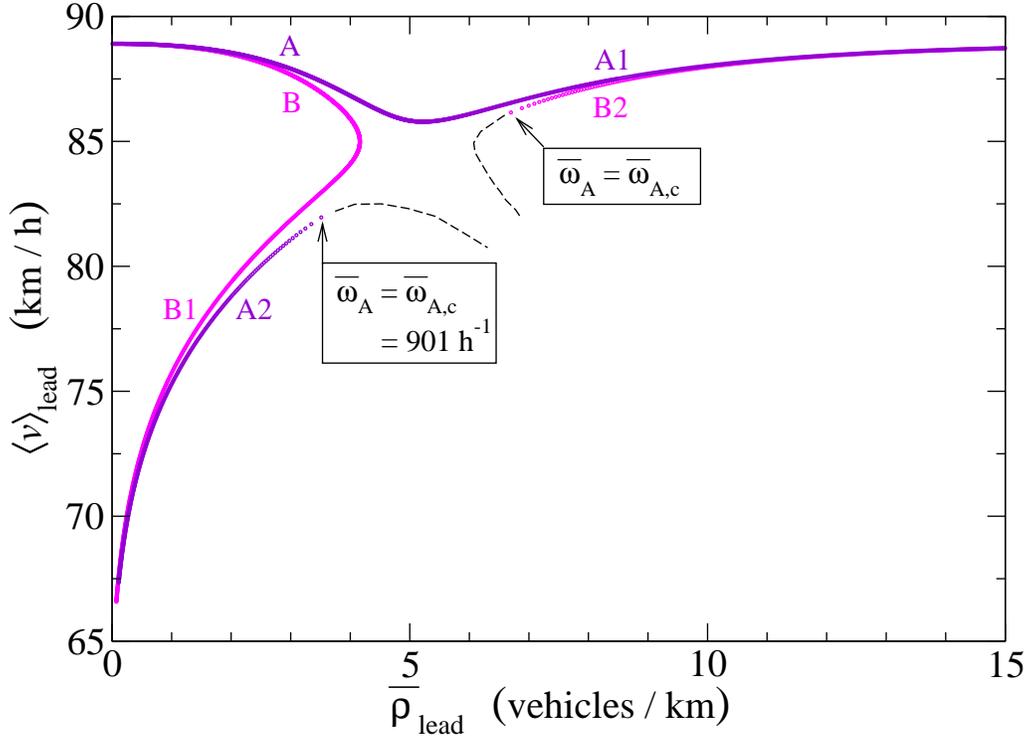}}
\end{center}
\caption{\small 
(OBC)
Locus of solutions of equations (\ref{statsymm})
in the asymmetric case with $\bomega_B=0.95\bomega_A$. Branches
A and B go over continuously into those marked A1 and B1, respectively.
At a critical value $\bomega_A=\bomega_{A,{\rm c}}\approx 901$ vehicles/hour 
a secondary solution appears with
two stable branches marked A2 and B2. This solution has $B$ as the fast
lane and $A$ as the slow one. The dashed lines schematically represent
unstable parts of the secondary branches.}
\label{asymXY}
\end{figure}
%%%%%%%%%%%%%%%%%%%%%%%%%%%%%%%%
%%%%%%%%%%%%%%%%%%%%%%%%%%%%%%%%

%%%%%%%%%%%%%%%%%%%%%%%%%%%%%%%%
%%%%%%%%%%%%%%%%%%%%%%%%%%%%%%%%
\begin{figure}
\begin{center}
\scalebox{.55}
%{\includegraphics{XY113c.eps}}
{\includegraphics{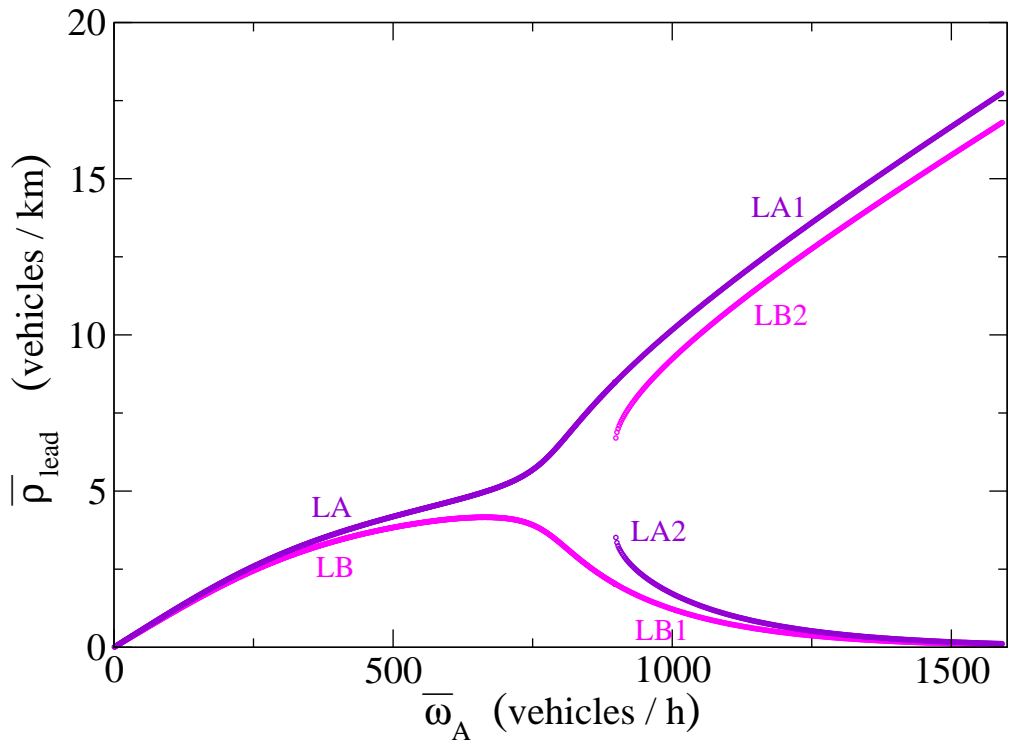}}
\end{center}
\caption{\small 
The leading vehicle density
  $\brho_{\rm lead}$ corresponding to figure \ref{asymXY}.
At $\bomega_A=\bomega_{A,{\rm c}}=901$ vehicles/hour
two new branches appear, marked here LA2 and LB2. 
The unstable branches are no longer shown.}
\label{asymdens}
\end{figure}
%%%%%%%%%%%%%%%%%%%%%%%%%%%%%%%%
%%%%%%%%%%%%%%%%%%%%%%%%%%%%%%%%

%%%%%%%%%%%%%%%%%%%%%%%%%%%%%%%%
%%%%%%%%%%%%%%%%%%%%%%%%%%%%%%%%
\begin{figure}
\begin{center}
\scalebox{.55}
%{\includegraphics{XY112c.eps}}
{\includegraphics{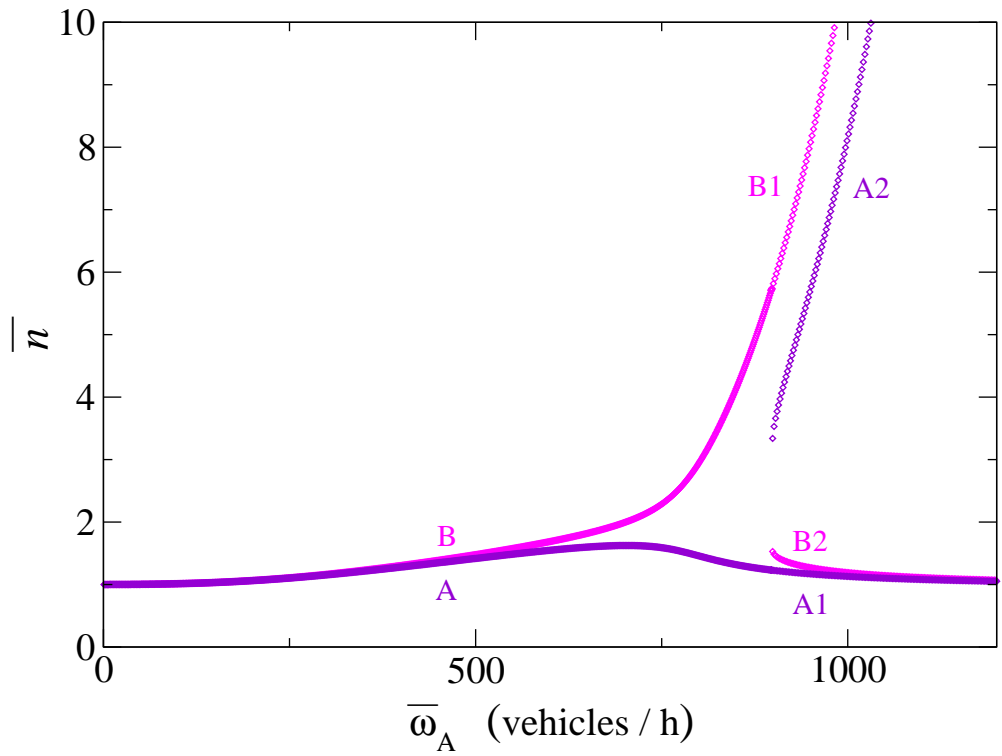}}
\end{center}
\caption{\small 
Average platoon length $\bar{n}$
  corresponding to figure \ref{asymXY}. 
The unstable branches are not shown.}
\label{asymn}
\end{figure}
%%%%%%%%%%%%%%%%%%%%%%%%%%%%%%%%
%%%%%%%%%%%%%%%%%%%%%%%%%%%%%%%%

%%%%%%%%%%%%%%%%%%%%%%%%%%%%%%%%%%%%%%%%%%%%%%%%%%%%%%%%%%%%%%%%%%%%%%%%%%%%%%

\subsection{Symmetric lanes with PBC}
\label{secpbc}

It is worthwhile also to see what happens when the open
boundary conditions (OBC) are replaced by periodic boundary conditions (PBC).
For one thing, PBC may be easier in simulations of this and other models.
In the case of PBC the density $\rho(v)\equiv\brho R(v)$ is given
and the total vehicle density $\brho$ is the control
parameter. We have studied the example where $R(v)$ is the same truncated
Gaussian, discretized on $N+1$ points $v_i$, that was used in section
\ref{secexamples}.

The flux $\bomega$ is now a derived quantity. We may decompose it
as the sum of the fluxes 
$\bomega_{\rm lead}$ and $\bomega_{\rm foll}$
of leading and following vehicles, respectively. 
In figure \ref{symrhodens} we have plotted $\bomega$ and 
$\bomega_{\rm lead}$ as functions of $\brho$. Both bifurcate at a
critical point $\brho_{\rm c}=9.02$ km$^{-1}$.

The behavior of the densities $\brho_{\rm lead}$ and $\brho_{\rm foll}$
may also be found as a function of $\brho$;
it is very similar to the case of OBC, but with the difference that
their sum $\brho$ is now a nonbifurcating straight line of unit slope.
It thus appears that OBC and PBC lead to the same qualitative
pictures. 
The qualitative features again remain unchanged as one varies the
distribution $R(v)$.

%%%%%%%%%%%%%%%%%%%%%%%%%%%%%%%%%%%%%%%%%%%%%%%%%%%%%%%%%%%%%%%%%%%%%%%%%%%%%%

\subsection{Non-symmetric lanes with OBC}
\label{secnonsymmetric}

There are many ways to break the symmetry between the two
lanes. We have chosen to investigate the open-ended problem with
$\bomega_A\neq\bomega_B$, all other things in the two lanes
being identical. In particular, $P_A(v)=P_B(v)=P(v)$ is again the Gaussian
distribution of equations (\ref{truncGauss})-(\ref{examplenum}).
We have taken a small symmetry breaking represented by
$\bomega_B=0.95\,\bomega_A$ and solved again equations (\ref{})
for $X_A, Y_A, X_B,$ and $Y_B$.
Instead of figure \ref{symXY} we now obtain figure \ref{asymXY}, in which
again the dashed lines represent (only schematically) branches of unstable 
solutions. 
Correspondingly, instead of the curves marked L, LA, and LB of figure
\ref{symdens} we now obtain the curves in figure \ref{asymdens}.
For low values of $\bomega_A$ (and concomitantly $\bomega_B$)
the platoon densities in the two lanes 
(the curves marked LA and LB) increase
linearly with $\bomega_A$, their slopes having a ratio 1\,:\,0.95. 
When $\bomega_A$ increases, the critical point is avoided
and the platoon density of lane $A$ (lane $B$) evolves continuously
to the values represented by the branch marked LA1 (LB1).
However, there is now a new critical value of
$\bomega_A=\bomega_{A,{\rm c}}=$901\,h$^{-1}$
at which an inversion of the densities becomes possible.
The two lanes are then on the `secondary' branches LA2 and LB2.

We are in the presence of a stationary state problem for which
there is no obvious way to define a free energy
functional, although this may not be impossible. 
However, it seems reasonable to consider this set
of secondary branches as `metastable' 
(in the sense of having a higher free energy)
with respect to the main branches.

The curves for the total and the following vehicle densities
(not plotted in figure \ref{asymdens}) have analogous behavior.
In figure \ref{asymn} we have plotted the corresponding average platoon
lengths. 

%%%%%%%%%%%%%%%%%%%%%%%%%%%%%%%%%%%%%%%%%%%%%%%%%%%%%%%%%%%%%%%%%%%%%%%%%%%%%%

\subsection{Upper density limit}
\label{secvaliditynum}

As we have seen in section \ref{secvalidity}, there is a high density regime
where the present model ceases to be valid. The condition for validity
depends on the external parameter $d_{\rm interveh}$, which represents
the typical distance between two consecutive vehicles in a platoon.
For the example of equations (\ref{example}) we may estimate
$d_{\rm interveh}=25$\,m, which when used in equation
(\ref{validity2}) gives 
\beq
\brho \ll 40\mbox{ vehicles/km}.
\label{valnum}
\eeq
As we have seen, the principal phenomenon investigated in this work,
namely the spontaneous symmetry breaking between two traffic lanes,
occurs at a density of $\brho\approx 5$ vehicles/km. Our conclusions
concerning this phenomenon are therefore compatible
with the simplifications of zero vehicle length and zero platoon
length that we used.

%%%%%%%%%%%%%%%%%%%%%%%%%%%%%%%%%%%%%%%%%%%%%%%%%%%%%%%%%%%%%%%%%%%%%%%%%%%%%%

\section{Mathematical remarks}
\label{secmathematical}

In this penultimate section we collect some comments on the
calculations presented at various stages of this work
with the purpose of clarifying their precise mathematical status.

The one-lane model is a statistical ensemble of geometrical constructs
of the type represented in figure \ref{figprim5}.
This ensemble is defined
by (i) the construction rules of sections \ref{secintroduction} and 
\ref{secsinglelane}; and (ii)
as an input, the statistics of the intersections of
the vehicle trajectories with one of the boundaries of the space-time domain,
either the $t$ or the $x$ axis.
One of the simplest ways of realizing (ii) would be to assume Poisson
statistics for the intersection points and to assign to the vehicles
velocities drawn independently 
from a given distribution, whether $P(v)$ or $R(v)$.
More generally, however, one could on the boundary
specify arbitrary correlations
between the positions and the velocities of the vehicles.

To develop the theory we have imposed,
in addition to $P(v)$ or $R(v)$, 
only the average flux $\bomega$ or average density
$\brho$, but have not specified any boundary correlations.
Our theory is therefore essentially a set of relations between 
{\it average\,} densities, fluxes, velocities, {\it etc.}

In section \ref{secnaturaleffective} 
the existence of a stationary state and a
well-defined function $\phi(v)$ is strictly speaking an assumption,
however natural it may be.
We leave open the possibility that the stationary state
correlations (but not the averages) 
depend on the unspecified boundary correlations.

On only two occasions did the stationary state
correlations between the vehicles play a role
and was an extra assumption about them needed.
The first occasion was when we discussed
the statistics of platoon lengths in section \ref{secplatoons};
this calculation was accessory and had no bearing on the rest of the paper.
The second occasion was in section \ref{sectau} and
appendix \ref{secapptau},
when we derived the function $F$ that connects the queuing times
in one lane to the traffic density in the other one; 
although this connection is the very essence of the two-lane model, 
we expect that different assumptions about the correlations
would lead to a qualitatively similar function $F$.
Nevertheless, in the single-lane model and for fully specified
boundary statistics,
it is a well-defined and still open question to determine the statistics of
the platoon arrival times at a fixed observation point. 
We have not attempted in this work to answer that question
and have simply presented the Poisson process nature
of the arrival times as an additional hypothesis.

%%%%%%%%%%%%%%%%%%%%%%%%%%%%%%%%%%%%%%%%%%%%%%%%%%%%%%%%%%%%%%%%%%%%%%%%%%%%%%

\section{Summary and conclusion}
\label{secconclusion}

In this paper we have proceeded in two steps.

First we defined a geometric traffic model of flux $\bomega_A$ on a single
lane  (`lane $A$'). The model includes the 
possibility for fast vehicles to overtake slower ones and 
is attractive because of its simplicity.
Vehicle trajectories are straight lines except that
overtaking vehicles incur time delays 
that follow from a given `queuing function' $\tau_A(v')$,
where $v'$ is the velocity of the vehicle being overtaken. 
The single-lane model exhibits formation of platoons of
vehicles. The platoon
density $\brho_{\rm lead}$ as compared to the total vehicle density $\brho$
has been one of our main quantities of interest.
We were able to relate analytically the platoon's statistics to
the flux intensity.

Secondly, and on the basis of additional hypotheses,
we expressed the queuing function $\tau_A$ governing the lane $A$ vehicles
in terms of the traffic conditions in the opposite lane, $B$,
and {\it vice versa.}
This resulted in a set of coupled two-lane equations.
Although the two lanes $A$ and $B$ 
have individually been modeled microscopically -- that is, in terms of the
trajectories of individual vehicles --,
the two-lane equations connect densities and average velocities in the
two lanes but no longer have a microscopic interpretation.
We solved the two-lane equations numerically for a characteristic
example of traffic flow
with {\it a priori\,} identical traffic conditions
on the two lanes ($\bomega_A=\bomega_B=\bomega$).
It was found that above a critical traffic flux, $\bomega>\bomega_{\rm c}$,
(or above a critical traffic density) 
the symmetry between the two lanes is spontaneously broken.
One lane has fast traffic with mostly $1$-platoons (= single vehicles), 
whereas the other lane has slow traffic with almost all vehicles
queuing in long platoons.
Asymmetric conditions $(\bomega_A\neq\bomega_B)$ have been briefly
investigated, as well as the case where not the fluxes $\omega_{A,B}$, but the
spatial vehicle densities $\brho_{A,B}$ are the control parameters.

We argued that the model considered in this work
is valid for low and moderate traffic densities.
A more elaborate version, appropriately defined and
including the parameter $d_{\rm interveh}$ (= the typical intervehicle
distance in a platoon),
would be able to describe also the very high density regime and 
the phenomenon of jamming of one or both lanes.
However, the spontaneous symmetry breaking reported in this paper
occurs in the range of densities for which the present model is valid.

It has not been our purpose to be exhaustive.
Several questions of interest have been left aside here 
but can be fairly straightforwardly
treated in the same model.
One of these concerns cases in which the velocity distribution has a 
tail that goes down to zero.
Another one is the incorporation,
along the road section under investigation,
of entrance or exit ramps with prescribed vehicle fluxes.
A less simple extension would be to include
time-varying boundary conditions as may be due,
for example, to traffic lights at the entrance of the road section.
We have considered only the stationary state;
studying the {\it relaxation\,} of an initial state
towards stationarity would allow, among other things, a comparison
with the scaling laws of reference \cite{BenNaimetal94}.

Analytically various limit cases could be pursued further.
In the weakly interacting limit an expansion of all 
relationships of this work in powers of $\bomega\tau_A$ and $\bomega\tau_B$
is certainly possible.
In the limit where the variance $\sigma^2$ of the velocity
distribution becomes small,
the phase transition point moves to higher densities; an expansion
for small variance seems more complicated but nevertheless also possible.
We have not exhibited any exact solutions to our equations;
no doubt a few exist and we think some of these
may be of interest.

The lane-lane coupling that we have introduced
in the present study is of {\it mean-field type,}
since each lane $A$ vehicle feels only the average effect 
of all lane $B$ vehicles, and {\it vice versa.}
One should wonder if this approximation is justified and if in its
absence the symmetry breaking still exists.
We believe it does, the main argument being that each vehicle in one
lane encounters in the course of time all vehicles in the other lane.
This argument is admittedly heuristic. 
Some light is shed on its validity
in two ways. First, in separate work \cite{appert_h_s10b}
we found that the simulation
of a closely related microscopic two-lane model also presents this
symmetry breaking, 
at least on practically relevant scales of time and space.
Secondly, in the quite different context of magnetic
friction Hucht and coworkers \cite{Kadauetal08,Hucht09}
were interested in an Ising 
system on a ladder lattice whose two legs are in relative motion.
This model has essentially the same features as the present road traffic
model; an exact solution \cite{Hucht09} shows that there is a phase
transition with mean-field characteristics for infinite
relative velocity, which however gets rounded when the velocity
becomes finite.
We will leave further discussion of these issues to the future.

%%%%%%%%%%%%%%%%%%%%%%%%%%%%%%%%%%%%%%%%%%%%%%%%%%%%%%%%%%%%%%%%%%%%%%%%%%%%%%
%%%%%%%%%%%%%%%%%%%%%%%%%%%%%%%%%%%%%%%%%%%%%%%%%%%%%%%%%%%%%%%%%%%%%%%%%%%%%%
%%%%%%%%%%%%%%%%%%%%%%%%%%%%%%%%%%%%%%%%%%%%%%%%%%%%%%%%%%%%%%%%%%%%%%%%%%%%%%

\appendix

%%%%%%%%%%%%%%%%%%%%%%%%%%%%%%%%%%%%%%%%%%%%%%%%%%%%%%%%%%%%%%%%%%%%%%%%%%%%%%

\section{Solution of Eq.\,\ref{eqnphiv} for $\phi(v)$}
\label{secappendixchi}

In this appendix we solve equation (\ref{eqnphiv}) for $\phi(v)$.
We denote the numerator of its RHS by
\beq
g(v)=v+\int_{v_0}^v\!\dd v'\,\omega(v')\tau(v')(v-v').
\label{defgv}
\eeq
This quantity satisfies
\beq
g'(v)=1+\int_{v_0}^v\!\dd v'\,\omega(v')\tau(v'), \qquad 
g^{{\pp}}(v)=\omega(v)\tau(v).
\label{}
\eeq
Upon setting $\phi(v)=1/\chi(v)$ 
we find that $\chi(v)$ satisfies the linear equation
\beq
\chi(v)g(v)=1+\int_{v_0}^v\!\dd v'\,\chi(v')(v-v')\omega(v')\tau(v')
\label{eqnchi1}
\eeq
whence by differentiating twice we obtain
\beq
[\,\chi(v) g(v)\,]'=\int_{v_0}^v\!\dd v'\,\omega(v')\tau(v')\chi(v')
\label{chigd1}
\eeq
and
\beq
[\chi g]^{{\pp}}=g^{{\pp}}\chi.
\label{chigd2}
\eeq
Setting now $\psi(v)=\chi'(v)$ we can rewrite (\ref{chigd2}) as
\beq
[\log\psi]'=-2[\log g]'\,.
\label{logpsid}
\eeq
Two successive integrations applied to (\ref{logpsid}) yield
\beq
\chi'(v)=\psi(v)=\frac{C_1}{g^2(v)}
\label{respsi}
\eeq
and
\beq
\frac{1}{\phi(v)}=\chi(v)=C_2+C_1\int_{v_0}^v\!\dd v'\,\frac{1}{g^2(v')}\,,
\label{reschi1}
\eeq
where $C_1$ and $C_2$ are constants of integration.
In order to determine these constants we first observe that
(\ref{reschi1}) leads to $1/\phi(v_0)=\chi(v_0)=C_2$.
Since for the reasons exposed in section \ref{secintroduction} we have
$\phi(v_0)=v_0$,
it follows that $C_2=1/v_0$.
Next we evaluate (\ref{chigd1}) for $v=v_0$. Using that $g(v_0)=v_0$
and $g'(v_0)=1$ this leads us to $v_0\chi'(v_0)=-\chi(v_0)$ whence
$\chi'(v_0)=-1/v_0^2$. Upon comparing this last result to (\ref{respsi})
evaluated at $v=v_0$ we see that $C_1=-1$. 
Substitution of the values of $C_1$ and $C_2$ in (\ref{reschi1})
yields the solution (\ref{solchiv}) given in the main text.
It is easily verified that for $\omega(v)=0$ equation (\ref{solchiv})
reduces to $\phi(v)=v$, as had to be the case.

%%%%%%%%%%%%%%%%%%%%%%%%%%%%%%%%%%%%%%%%%%%%%%%%%%%%%%%%%%%%%%%%%%%%%%%%%%%

\section{An analytic example of functions 
$\phi(v)$ and $\rho_{\rm lead}(v)$}
\label{secappexactsol}

%%%%%%%%%%%%%%%%%%%%%%%%%%%%%%%%%%%%%%%%%%%%%%%%%%%%%%%%%%%%%%%%%%%%%%%%%%%

\subsection{Analytic example of $\phi(v)$}
\label{secexamplephi}

Let $\omega(v)=\bomega P(v)$ be
a block distribution of the natural velocities,
\beq
\omega(v)=\left\{
\begin{array}{ll}
\dfrac{\bomega}{\vm-v_0} \phantom{XXX}& v_0<v<\vm, \nonumber\\[4mm]
0                                      & \mbox{else},
\end{array}
\right.
\label{defexactomega}
\eeq
let $\tau(v)=\btau$, and define the dimensionless parameter 
$\lambda=\bomega\btau$. 
Integrating according to equation (\ref{defgv}) 
we first find
\beq
g(v) = v + \lambda \frac{(v-v_0)^2}{\vm-v_0}.
\label{exexactgv}
\eeq
To abbreviate the notation we set
\bea
\begin{array}{ll}
k(u) = v_0 + u + cu^2, \phantom{XXXX}& u = v-v_0\,, \\
k'(u) = 1 + 2cu,        & c = \lambda/(\vm-v_0)\,, \\
& \Delta = -1 + {2v_0\lambda}/(\vm-v_0)\,.
\end{array}
\label{defapp}
\eea
Next we find from equation (\ref{reschi1}) that 
\beq
\frac{1}{\phi(v)} = \frac{1}{v_0} + I(v-v_0)-I(0),
\label{exexactphi}
\eeq
where $I(u)$ is the indefinite integral
\bea
I(u) &=& \int\!\frac{\dd u}{k^{\,2}(u)} \nonumber\\[2mm]
&=& \frac{k'(u)}{\Delta k(u)} 
    -\frac{4c}{\Delta} 
\left\{
\begin{array}{lll}
\phantom{-}\dfrac{1}{\sqrt{\Delta}}\,
\mbox{arccot} \dfrac{k'(u)}{\sqrt{ \Delta}},\qquad&\Delta>0,\,\,\\[6mm]
\dfrac{1}{\sqrt{-\Delta}}\,
\mbox{artanh} \dfrac{k'(u)}{\sqrt{-\Delta}},      &\Delta<0,\,\,\, 
              \dfrac{k'(u)}{\sqrt{-\Delta}}<1,\\[6mm] 
\dfrac{1}{\sqrt{-\Delta}}\,
\mbox{arcoth} \dfrac{k'(u)}{\sqrt{-\Delta}},      &\Delta<0,\,\,\, 
              \dfrac{k'(u)}{\sqrt{-\Delta}}>1. \\
\end{array} 
\right.
\nonumber\\
&&
\label{resexactI}
\eea
Depending on the values of the parameters and the variables 
of the problem, all three cases of this equation may occur.
The desired function $\phi(v)$ now follows by substitution of  
(\ref{resexactI}) in (\ref{exexactphi}).

%%%%%%%%%%%%%%%%%%%%%%%%%%%%%%%%%%%%%%%%%%%%%%%%%%%%%%%%%%%%%%%%%%%%%%%%%%%%

\subsection{Limit of small $\lambda$}
\label{secsmalllambda}

The above result becomes easier to visualize in the limit of small
$\lambda$. In that limit the `arcoth' applies in equation
(\ref{resexactI}). 
After a certain amount of algebra one obtains
\beq
\phi(v) = v +\frac{\lambda}{\vm-v_0}\left[
(3v-v_0)(v-v_0)-2v^2\log\frac{v}{v_0}
\right] + {\cal O}(\lambda^2),
\label{explambda}
\eeq
in which the expression in brackets may be shown to be negative for
all $v>v_0$, as had to be the case because of $\phi(v)<v$.
Expanding the coefficient of the term linear in $\lambda$
near the lower limit $v_0$ of the velocity distribution
we get
\beq
\phi(v_0+u)= v_0 + u -\lambda\left[\frac{u^3}{3v_0(\vm-v_0)} 
+{\cal O}(u^4)\right] + {\cal O}(\lambda^2).
\label{explambdau}
\eeq

Let $\la w\ra_\lambda=\la\phi(v)\ra_\lambda$ 
denote the effective velocity $w$ averaged with repect to
$P(v)$ at a given $\lambda=\bomega\btau$.
Then we find from (\ref{defexactomega}) together with (\ref{explambdau})
\beq
\la w\ra_\lambda = \la v\ra_0\,+\,\frac{\lambda}{2(\vm-v_0)}
\left[ \tfrac{11}{9}\vm^3 -2\vm^2v_0 + \vm v_0^2 -\tfrac{2}{9}v_0^3
-\tfrac{2}{3}\vm^3\log\frac{\vm}{v_0} \right] + {\cal O}(\lambda^2)
\label{avvlambda}
\eeq
in which $\la w\ra_0=\la v\ra_0=\tfrac{1}{2}(\vm-v_0)$ 
is just the average natural velocity.
In the limit of a narrow distribution, $\vm-v_0\to 0$, this becomes
\beq
\la w\ra_\lambda = \la v\ra_0\,-\,\frac{\lambda}{12v_0}
\left[ (\vm-v_0)^2 + {\cal O}\Big( (\vm-v_0)^3 \Big) \right]
+ {\cal O}(\lambda^3).
\label{avvlambdanarrow}
\eeq
The factor $(\vm-v_0)^2$ shows that the reduction of the average
effective velocity is proportional to the variance of the distribution.

%%%%%%%%%%%%%%%%%%%%%%%%%%%%%%%%%%%%%%%%%%%%%%%%%%%%%%%%%%%%%%%%%%%%%%%%%%%

\subsection{Analytic example of $\rho_{\rm lead}(v)$}
\label{secappdiscrrho}

The platoon density $\rho_{\rm lead}(v)$ corresponding 
to the present example is easiest to obtain from equation
(\ref{exrholeadsim}) [which is also (\ref{rholeadd})-(\ref{defKv})], 
from which $\phi(v)$ has been eliminated.
Upon observing that $K(v)=k(v-v_0)$ we get straightforwardly from
(\ref{rholeadd}) the simple expression 
\beq
\rho_{\rm lead}(v)=\frac{\bomega}{(\vm-v_0)+\tfrac{1}{2}\lambda(v-v_0)^2}\,,
\label{resexactrholead}
\eeq
from which many further analytic results may be derived by means of
the various relations of this work.

%%%%%%%%%%%%%%%%%%%%%%%%%%%%%%%%%%%%%%%%%%%%%%%%%%%%%%%%%%%%%%%%%%%%%%%%%%%

\section{Discretized equations}
\label{secappdiscr}

One may discretize the velocity axis by 
setting $v_i=v_0+i\Delta v$, where 
$\Delta v$ is a fixed increment. We let $i$ run from $0$ to some large
value $N$ that may later on be set equal to infinity.
The discretized version of the equations is used
for the numerical solutions discussed in this work.
It is, however, more than just a way of approximating the continuum
limit, since it describes
a traffic model having only a discrete set of velocities.

{\it Open boundary conditions.\,\,}
Let us now take for $\omega(v)=\bomega P(v)$ the discrete distribution
\beq
\omega(v)=\sum_{i=0}^N \omega_i\, \delta(v-v_i), \qquad 
\sum_{i=0}^N \omega_i = \bomega,
\label{defPdiscr}
\eeq
and set
\beq
\tau_i=\tau(v_i).
\label{deftaudiscr}
\eeq
Upon substituting (\ref{defPdiscr}) and (\ref{deftaudiscr}) 
in (\ref{eqnphiv}) and abbreviating
$\phi_i=\phi(v_i)$ we get for $\phi_j$ the equation
%%%[*** en fait, on peut refaire le calcul en d\'etail, ce qui est plus s\^ur]
\beq
\phi_j=\frac{v_j+\Delta v \sum_{i=1}^{j-1}(j-i)\omega_i\tau_i}
{1+\Delta v \sum_{i=1}^{j-1}(j-i)\omega_i\tau_i\phi_i^{-1}}\,,
\qquad j=1,\ldots,N,
\label{eqnphicdis}
\eeq
in which $\tau_i\equiv\tau(v_i)$ and
where the right hand member depends only on $\phi_0,\ldots,\phi_{j-1}$.
Given that $\phi_0=v_0$ we may obtain from
this equation successively the numerical values of
$\phi_1,\phi_2,\ldots,\phi_N.$ One can easily see that they are 
increasing with $j$. 
Equation (\ref{eqnphicdis}) is the discrete counterpart of (\ref{solchiv}).
Its explicit solution $\phi_j\equiv\phi(v_j)$ is
\beq
\phi_j = \frac{1}{v_0}\,-\,\Delta v\sum_{i=1}^j\frac{1}{g_{i-1}g_i}
\label{solphivdis}
\eeq
with
\beq
g_j=v_j\,+\,\Delta v\sum_{i=0}^{j-1}(j-i)\omega_i\tau_i\,.
\label{defgdis}
\eeq
When the velocities have the discrete distribution (\ref{defPdiscr})
we may set 
\beq
\rho_{\rm lead}(v) = \sum_{j=0}^N \rho_{{\rm lead},j}\,\delta(v-v_j)
\label{defrholeaddiscr}
\eeq
and find that the discrete equivalent of (\ref{exrholeadsim}) is
\beq
\rho_{{\rm lead},j}= \omega_j g_j^{-1},
\qquad j=0,1,\ldots,N,
\label{exrholeadsimdiscr}
\eeq
where we use the notation of appendix \ref{secappdiscr}. 

{\it Periodic boundary conditions.\,\,}
In the case of cyclic boundary conditions the density $\rho(v)$
is the control function. Let us suppose it is of the discrete type
\beq
\rho(v)=\sum_{i=0}^N \rho_i\, \delta(v-v_i).
\label{defrhodiscr}
\eeq
Substitution of this expression in (\ref{exrholeadpbc})
yields again expression (\ref{eqnphicdis}) but with $\omega p_i$
replaced with $\rho_i\phi_i$.
Given that $\phi_0=v_0$ we may determine successively $\phi_1,\ldots,\phi_N$ 
in the same way as for open boundary conditions. 

Solving the equation explicitly yields
\beq
\phi_j = v_0 + \Delta v\sum_{i=1}^j\frac{1}{h_{i-1}h_{i}},
\qquad j=0,1,\ldots,N
\label{solphicdispbc}
\eeq
with
\beq
h_j = 1 + \Delta v \sum_{i=0}^{j-1}(j-i)\rho_i\tau_i\,,
\qquad j=0,1,\ldots,N.
\label{defhdis}
\eeq
These two equations are the discrete versions of (\ref{solphivpbc}) and 
(\ref{defhv}), respectively.

In the case of cyclic boundary conditions the expression for the density
$\rho(v,v')$ is again given by (\ref{deftrhominvvp}) but with $\omega$ 
replaced with $\rho$. Its discrete equivalent, with an obvious
definition of $\rho_{ij}$, is 
\beq
\rho_{ij} = \rho_i\rho_j\tau_j(\phi_i-\phi_j).
\label{defrhoij}
\eeq
Similarly, equation (\ref{exrholead}) has, in the case of cyclic boundary 
conditions, the discrete equivalent
\beq
\rho_{{\rm lead},j} = \rho_j\left[ 1 - 
\sum_{i=0}^{j-1}\rho_i\tau_i(\phi_j-\phi_i) \right].
\label{exrholeaddispbc}
\eeq
When (\ref{solphicdispbc}) is used in (\ref{exrholeaddispbc}) we obtain
after substantial rewriting the simplified result
\beq
\rho_{{\rm lead},j} = \rho_j h_j^{-1}
\label{exrholeaddissimpbc}
\eeq
with $h_j$ given by (\ref{defhdis}).
Equations (\ref{exrholeaddissimpbc}) and (\ref{defhdis}) 
constitute the discrete equivalent of (\ref{exrholeadsimpbc})
and (\ref{defhv}) combined.

%%%%%%%%%%%%%%%%%%%%%%%%%%%%%%%%%%%%%%%%%%%%%%%%%%%%%%%%%%%%%%%%%%%%%%%%%%%

\section{Rewriting Eq.\,(\ref{exrholead})}
\label{secArholead}

We consider here expression (\ref{exrholead}) for the platoon density
$\rho_{\rm lead}$,  
\beq
\rho_{\rm lead}(v) 
= \omega(v)
\left\{ \,\frac{1}{\phi(v)}\,-\,\int_{v_0}^v \!\dd v'
\left[\frac{1}{\phi(v')}-\frac{1}{\phi(v)} \right] \omega(v')\tau(v') \right\},
\label{rholeadb}
\eeq
and show how it
may be cast in a much simpler form. It is useful to abbreviate
\beq
K(v) = v\,+\,\int_{v_0}^{v}\!\dd v'\,(v-v')\omega(v')\tau(v').
\label{defKv}
\eeq
Hence 
\beq
\frac{1}{\phi(v)}= \frac{1}{v_0}\,-\,\int_{v_0}^v\!\frac{\dd v_1}{K^2(v_1)} 
\label{relphiK}
\eeq
and, for $v'<v$,
\beq
\frac{1}{\phi(v')} - \frac{1}{\phi(v)} = 
\int_{v'}^v\!\frac{\dd v_1}{K^2(v_1)}\,.
\label{reldiffphiK}
\eeq
Using (\ref{reldiffphiK}) in (\ref{rholeadb}) we obtain 
\bea
\rho_{\rm lead}(v) &=& \omega(v) \left\{ \frac{1}{\phi(v)}\,
-\,\int_{v_0}^v\!\dd v'\,\int_{v'}^v\!\dd v_1\,\frac{1}{K^2(v_1)} 
\omega(v')\tau(v')\right\}
\nonumber\\[2mm]
&=& \omega(v) \left\{ \frac{1}{\phi(v)}\,
-\,\int_{v_0}^v\!\dd v'\,\frac{1}{K^2(v')}\, 
\int_{v_0}^{v'}\!\dd v_2\,\omega(v_2)\tau(v_2)\right\},\phantom{XX}
\label{rholeadc}
\eea
where to arrive at the second line we performed an integration by parts.
When with the aid of (\ref{relphiK}) we eliminate the $1/\phi(v)$
on the right hand side of (\ref{rholeadc}) we obtain 
\bea
\rho_{\rm lead}(v) &=& \omega(v) \left\{ \frac{1}{v_0}\,
-\,\int_{v_0}^v\!\dd v'\,
\frac{1\,+\,\int_{v_0}^{v'}\!\dd v_2\,\omega(v_2)\tau(v_2)}{K^2(v')}\, 
\right\}
\nonumber\\[2mm]
&=& 
\omega(v) \left\{ \frac{1}{v_0}\,
+\,\int_{v_0}^v\!\dd v'\,
\frac{1}{K^2(v')}\frac{\dd K(v')}{\dd v'} 
\right\}
\nonumber\\[2mm] 
&=&
\frac{\omega(v)}{K(v)}\,,
\label{rholeadd}
\eea
where we used that $K(v_0)=v_0$.
Upon substituting for $K(v)$ in the last line of (\ref{rholeadd})
its definition (\ref{defKv}) one obtains the final result of this rewriting, 
equation (\ref{exrholeadsim}) of the main text.

%%%%%%%%%%%%%%%%%%%%%%%%%%%%%%%%%%%%%%%%%%%%%%%%%%%%%%%%%%%%%%%%%%%%%%%%%%%
%%%%%%%%%%%%%%%%%%%%%%%%%%%%%%%%%%%%%%%%%%%%%%%%%%%%%%%%%%%%%%%%%%%%%%%%%%%%

\section{An equation for $\phi(v)$ in the case of PBC}
\label{appsecphiPBC}

For the case of a single traffic lane with periodic boundary
conditions,
studied in section \ref{seccircular},
we wish to find an expression for $\phi(v)$ in 
terms of the control function $\rho(v)$.
The reasoning resembles the one of subsection \ref{seceqnphi}, 
but is still sufficiently 
different that we repeat the main steps here.
Let $T(v)$ be the time needed for a vehicle of natural velocity $v$ to cover a 
distance $L$, that is, to complete one full turn of the ring.
 Then
\beq
T(v) = \frac{L}{\phi(v)} = \frac{L}{w}\,.
\label{defTv}
\eeq
We consider a marked vehicle of natural velocity $v$ and hence of
effective velocity  
$w = \phi(v)$. Let $v'<v$ and $w'=\phi(v')$, so that $w'<w$.
Let $N(w,w')\dd w'$ denote the average number of vehicles
with velocities in the interval $[w',w'+\dd w']$ overtaken by the
marked vehicle. 
To find an explicit expression for this number, we notice that the total 
number of vehicles on the ring
having a velocity in the given $w'$ interval is $L\trho(w')\dd w'$.
Of these, the fraction overtaken by the marked vehicle is equal to
$1/L$ times the 
relative distance $(w-w')T(v)$ covered, that is,
\bea
N(w,w')\dd w' &=& \frac{(w-w')T(v)}{L} \times \trho(w')\dd w' \nonumber \\[2mm]
              &=& L \frac{w-w'}{w} \trho(w')\dd w'.
\label{exNwwp}
\eea
As before, the time lost in overtaking a vehicle of velocity $v'$
will be denoted by $\tau(v')$. Then,
due to its overtaking of vehicles with velocities in  $[w',w'+\dd w']$,  
the marked vehicle, while it cycles the ring once, will be queuing 
during a total time $\tilde{T}_{\rm foll}(w,w')$ that is given by
\beq
\tilde{T}_{\rm foll}(w,w') \dd w' = \tau(v')N(w,w')\dd w' 
\label{exTwwp}
\eeq
(in hybrid notation where we use both $v$ and $w$ variables).
The total distance $\tilde{d}_{\rm foll}(w,w')\dd w'$ 
that it covers during this lost time is
\beq
\tilde{d}_{\rm foll}(w,w') \dd w' = \tau(v')N(w,w')v'\dd w'. 
\label{exdwwp}
\eeq
Integrating now over all $w'$ we find that the total time loss 
$\tilde{T}_{\rm foll}(w)$ suffered by 
the marked vehicle during one cycle is
\beq
\tilde{T}_{\rm foll}(w) = \int_0^w\!\dd w'\, \tau(v')N(w,w').
\label{defTw}
\eeq
Similarly the total distance $\tilde{d}_{\rm foll}(w)$
covered during this lost time is
\beq
\tilde{d}_{\rm foll}(w) = \int_0^w\!\dd w'\, \tau(v')N(w,w')v'. 
\label{defdw}
\eeq
In the remaining time interval $T-\tilde{T}_{\rm foll}(w)$ the remaining 
distance $L-\tilde{d}_{\rm foll}(w)$ is covered at a velocity $v$. 
Hence we must have the key equation
\beq
v\big[ T-\tilde{T}_{\rm foll}(w) \big] = L-\tilde{d}_{\rm foll}(w).
\label{key1eq}
\eeq
We substitute (\ref{defTw}) and (\ref{defdw}) 
in this equation, using for $N(w,w')$ the explicit 
expression (\ref{exNwwp}). The result is the relation
\beq
v\left[ \frac{1}{w} - \int_{v_0}^w\! \dd w'\,\tau(v') \frac{w-w'}{w}\trho(w')
\right]
= 1-\int_{v_0}^w\! \dd w'\, \tau(v')\frac{w-w'}{w} v'\trho(w').
\label{key2eq}
\eeq
When we divide by $v$ and transform from $w$ and $w'$ to $v$ and $v'$
using (\ref{deftrhov}) we get
\beq
\frac{1}{\phi(v)}-\frac{1}{v} = \int_{v_0}^v\!\dd v'\,\tau(v')
\left(1-\frac{\phi(v')}{\phi(v)}\right)
\left(1-\frac{v'}{v}\right) \rho(v').
\label{key3eq}
\eeq
We can rewrite this as (\ref{key4eq}), which completes the proof and
indirectly confirms the validity of (\ref{extrhovPBC}) for the case of PBC.

%%%%%%%%%%%%%%%%%%%%%%%%%%%%%%%%%%%%%%%%%%%%%%%%%%%%%%%%%%%%%%%%%%%%%%%%%%%%%
%%%%%%%%%%%%%%%%%%%%%%%%%%%%%%%%%%%%%%%%%%%%%%%%%%%%%%%%%%%%%%%%%%%%%%%%%%%%%
%%%%%%%%%%%%%%%%%%%%%%%%%%%%%%%%%%%%%%%%%%%%%%%%%%%%%%%%%%%%%%%%%%%%%%%%%%%%%

\section{Relation between  $\tau_{A,B}(v)$  and  $\gamma_{A,B}(v)$}
\label{secapptau}

%%%%%%%%%%%%%%%%%%%%%%%%%%%%%%%%%%%%%%%%%%%%%%%%%%%%%%%%%%%%%%%%%%%%%%%%%%%%%%

\subsection{A problem in statistics}
\label{secstatistics}

The question of finding a relation between $\gamma_{A,B}(v)$ and $\tau_{A,B}(v)$
leads us to the following problem in statistics.
Point events take place at a rate $\gamma$.
We start observing them at an initial time $t=0$.
Let the stochastic variable $\htau$ denote the instant of time at which
the first time interval starts that has a length  
greater than $\tau_0$ and is free of events. 
The probability distribution $p(\htau)$ of $\htau$ will be of the form
\beq
p(\htau)=A\delta(\htau)+B(\htau),
\label{expt}
\eeq
where $A=\ee^{-\gamma\tau_0}$ is the probability that there is no event
in $[0,\tau_0]$ and where $B(\htau)\dd\htau$ is the probability that
the first eventless interval greater that $\tau_0$ starts in 
$[\htau,\htau+\dd\htau]$.
The calculation of $B(\htau)$ is possible but the expressions become
more complicated than is needed for our purpose.
We will therefore do something simpler,
but good enough for our purpose, 
namely replace $\htau$ by its {\it average\,} 
\beq
\tau = \int_0^\infty \!\,\htau\,p(\htau)\,\dd\htau.
\label{deftau}
\eeq
For dimensional reasons we can write
\beq
\tau=\tau_0 F(\gamma\tau_0).
\label{defF}
\eeq
Furthermore we should necessarily have
\beq
F(0)=0, \qquad F(\infty)=\infty.
\label{propF}
\eeq
It is not difficult to find an explicit expression for $\tau$ as defined by
(\ref{deftau}). 

%%%%%%%%%%%%%%%%%%%%%%%%%%%%%%%%%%%%%%%%%%%%%%%%%%%%%%%%%%%%%%%%%%%%%%%%%%%%%%

\subsection{The expression for $\tau_{A,B}(v)$}
\label{secextaukv}

Let $p_0$ be the probability that there is no event in $[0,\tau_0]$, and let
$p_n$ be the probability that the first time interval of length greater than
$\tau_0$ comes along after the $n$th event, for $n=1,2,\ldots$.
Then
\beq
p_n=(1-\ee^{-\gamma\tau_0})^n \,\ee^{-\gamma\tau_0}, \qquad n=0,1,2,\ldots
\label{expn}
\eeq
Using (\ref{deftau}) we get
\beq
\tau = \sum_{n=0}^\infty n\tau_{\rm av}^<\, p_n
\label{extau}
\eeq
where $\tau_{\rm av}^<$ is the average length of a time interval given that it
is shorter than $\tau_0$. 
An elementary calculation gives
\bea
\tau_{\rm av}^<&=& \int_0^{\tau_0}\!\dd t\,\frac{\gamma t\,\ee^{-\gamma t}}
{1-\ee^{-\gamma\tau_0}} \\[2mm]
&=& \frac{ 1-\ee^{-\gamma\tau_0}-\gamma\tau_0\ee^{-\gamma\tau_0} }
{ \gamma( 1-\ee^{-\gamma\tau_0} ) }
\label{restau1}
\eea
When (\ref{restau1}) is substituted in (\ref{extau}) we get
\bea
\tau &=& \sum_{n=0}^\infty n\tau_{\rm av}^<\,p_n \nonumber\\[2mm]
&=& \tau_{\rm av}^<(\ee^{\gamma\tau_0} - 1) \nonumber\\[2mm]
&=& [ \ee^{\gamma\tau_0} -1-\gamma\tau_0 ]/\gamma,
\label{calctau}
\eea
which is (\ref{restau}). It is indeed of the form (\ref{defF})-(\ref{propF})
with $F(x)=(\ee^x-1-x)/x$.

%%%%%%%%%%%%%%%%%%%%%%%%%%%%%%%%%%%%%%%%%%%%%%%%%%%%%%%%%%%%%%%%%%%%%%%%%%%%%%

\appendix

\end{document}